\documentclass[journal]{IEEEtran}

\usepackage[utf8]{inputenc} 
\usepackage[T1]{fontenc}    
\usepackage{hyperref}       
\usepackage{url}            
\usepackage{booktabs}       
\usepackage{amsfonts}       
\usepackage{nicefrac}       
\usepackage{microtype}      
\usepackage{lipsum}		
\usepackage{graphicx}
\usepackage{amsmath}
\usepackage{amssymb}
\usepackage[english]{babel}
\usepackage{float}
\usepackage{multirow}
\usepackage{adjustbox}
\usepackage{mathtools}
\usepackage[dvipsnames]{xcolor}
\usepackage[
backend=biber,
sorting=ynt
]{biblatex}
\usepackage{subcaption}
\usepackage{adjustbox}

\title{Bibliography management: \texttt{biblatex} package}
\author{Share\LaTeX}
\date{ }\usepackage{xcolor}

\newcommand{\comment}[1]{}

\usepackage{scalerel,stackengine}
\stackMath
\newcommand\reallywidehat[1]{%
\savestack{\tmpbox}{\stretchto{%
  \scaleto{%
    \scalerel*[\widthof{\ensuremath{#1}}]{\kern-.6pt\bigwedge\kern-.6pt}%
    {\rule[-\textheight/2]{1ex}{\textheight}}
  }{\textheight}%
}{0.5ex}}%
\stackon[1pt]{#1}{\tmpbox}%
}
\begin{document}

\title{Asymmetric Coding for Ultrahigh Throughput Encoding (ACUTE)}

\author{Seyed Mehdi Ayyoubzadeh, Xiaolin Wu~\IEEEmembership{Fellow}, Xi Zhang}

\maketitle
\begin{abstract}
Thanks to steady advancement of the imaging sensor technology, modern cameras can achieve very high precision in space and time. However, increasing the video frame rate while maintaining high spatial resolution is an arduous task because of the limited memory bandwidth to handle the huge data flow from the sensor array.  The compression algorithms for high speed cameras must have an encoder
whose throughput at least matches camera's data rate. Also, the encoder needs to be simple enough for implementation in camera's hardware pipeline.  On the other hand, the decoder is not constrained by the limited on-camera computational resources and battery life; it can work off line and use powerful computers, such as GPUs, to reconstruct high speed videos.
In this research, we propose an Asymmetric Coding scheme for Ultrahigh Throughput Encoding (ACUTE).  ACUTE consists of a simple and fast encoder that can compress raw sensor data as fast as they are read out.  
In contrast to the light-duty encoder, the ACUTE decoder is a heavy-duty deep decompression CNN model that can achieve good rate-distortion performance.  The key technical innovation is gradient coding by fast 2D lattice vector quantization at the encoder and optimized deep dequantization and super-resolution at the decoder.
\end{abstract}


\IEEEpeerreviewmaketitle

\section{Introduction}
\label{section:Introduction}
Thanks to ever increasing sophistication and capabilities of imaging technologies, professionals of many technical fields can now acquire images of very high resolutions simultaneously in spatial, spectral and temporal domains.  One example is the wide use of high-speed cameras that can reach peak frame rate 1000 Hz and above; they find applications in safety studies (e.g., crash tests of automobiles), studies of high speed phenomena, aerospace, manufacturing and production, entertainment, etc.  High speed in vivo imaging and ultrafast functional medical imaging are other examples.
In remote sensing, hyper-spectral images need to offer high resolutions in both spatial and spectral domains.
The above high-end imaging processes all require an ultrahigh data throughput. The total generated data volume per unit time is the product of the three resolutions in space, time and wavelength.  As a result, the limiting factor for the total achievable precision over all domains is the memory bandwidth of the imaging device.  This is why high speed cameras are compelled to sacrifice spatial resolution for higher temporal resolution.

In order to push the envelop of achievable precision in all imaged dimensions, one has to overcome the memory bandwidth bottleneck and hence comes the necessity of image compression.
But unlike conventional image compression tasks, multidimensional high-resolution image compression faces a special challenge: the encoder throughput
has to be high enough to match or exceed the memory bandwidth; it is the lower of the two speed specifications that determines how high a data rate can be sustained for a long duration of continuous shooting of high fidelity multidimensional cameras.
A case in point is compression for ultrahigh speed video cameras.  Conventional image and video compression methods, such as H.264/H.265 \cite{wiegand2003overview, sullivan2012overview}, JPEG  \cite{wallace1992thejpeg}, JPEG 2000, are not suited for the task. Although offering high compression, their encoding throughput is not high enough to sustain continuous operations of high speed cameras.
\begin{figure}[H]
  \centering
  \includegraphics[width=0.9\linewidth]{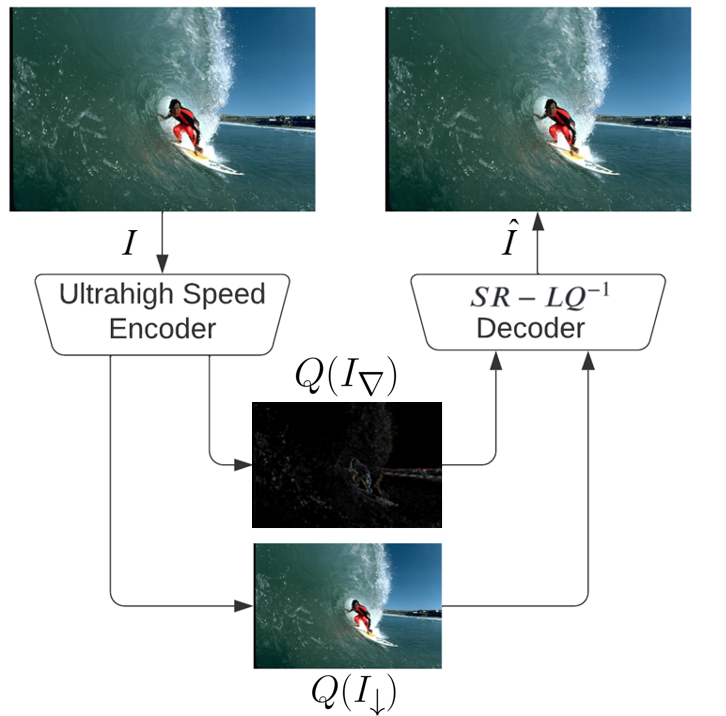}
  \caption{\textbf{Schematic diagram of the ACUTE system}. The encoder is exceptionally lightweight to achieve very high throughput. The compressed data consists of a downsampled image $I_\downarrow$ and a lattice quantized gradient image $I_\nabla$.  Decompression is done by a dual task CNN model (SR-LQ$^{-1}$) that performs soft gradient dequantization and superresolution.}
  \label{fig:outline}
\end{figure}
The encoders of these compression methods all require a 2D signal transform (either DCT or wavelet), transform coefficient quantization, entropy coding (either Huffman or arithmetic coding), and in the case of H.264/H.265 motion compensation.
Even if the 2D signal transformation can be implemented fast enough to keep up with the required throughput, it still must buffer many rows of the pixels in the image (8 in the case of DCT) in the readout process, generating a serious blockage of the data pipeline of high speed cameras.
Apparently, none of existing image compression methods is designed to handle the ultrahigh data rate of video cameras running at 1000 Hz or above.

In theory, compressive sensing (CS) is a possible solution to the problem.  CS can greatly reduce the sampling rate and hence generate a much smaller amount of data in the first place.
But none of CS-based compression techniques can match the rate-distortion performance of conventional image coding methods. In \cite{wang2012binned}, Wang et al.\ tried to mitigate the issue by a more efficient quantization technique; however, their encoder still cannot match the high throughput of high speed cameras.

Another strategy to reduce the encoder complexity without sacrificing rate-distortion performance much is distributed source coding \cite{dragotti2009distributed}.  The idea is not to aim at full removal of high-order statistical redundancies at the light encoder, but rather exploit such redundancies to make coding gains by the LZ-type algorithm \cite{ziv1977universal} at the heavy decoder.  In fact, distributed video coding (DVC) has been thoroughly investigated.  Compared with conventional hybrid video coding paradigm, DVC escapes the expensive motion estimation step and hence greatly reduces the encoder complexity \cite{girod2005distributed}.  But the DVC encoder is still too expensive to meet the stringent throughput requirement of high speed cameras.

To our best knowledge, the highest encoder throughput of digital cameras is achieved by what we propose in this paper, called the strategy of Asymmetric Coding for Ultrahigh Throughput Encoding (ACUTE).  As outlined in Figure \ref{fig:outline}, the ACUTE image compression system employs, based on the principle of distributed source coding, a light-duty, hardware-friendly encoder and a coupled heavy-duty decoder.   In an ACUTE system, the encoder complexity is reduced to minimum by forgoing prediction or transform, and by using fixed length code without entropy coding.  The compression is performed by simple downsampling by a factor of 2 and storing a small piece of prior information on the original.  The prior information is used to assist the decoder to solve an essentially image superresolution problem.  Here we stress that the decoder of ACUTE should be expected to outperform SR methods, because the encoder knows the exact HR ground truth image, which is unknown for the conventional SR task, and sends the decoder the prior information
on the original HR image.

The rest of the paper is structured as follows. We propose the design procedure for a high-speed encoder. Then we introduce our deep learning-based decoder (SR-LQ$^{-1}$), followed by the experiments that show the effectiveness of ACUTE. Finally, we do the ablation study to demonstrate the effectiveness of each designed module.



\section{Deep Learning based ACUTE Paradigm}

In the ACUTE system design, the computation burdens of achieving high rate-distortion performance are shifted from encoder to decoder.
No variable-length entropy coding greatly simplifies the code stream organization and parsing at the hardware level.  Such a maximally streamlined encoder can be integrated into an image sensor, and so it can achieve a coding speed as fast as sensor array readout speed, while still achieving compression ratio from 2:1 to 3:1.  If the ACUTE decoder can recover the original image in perceptually lossless quality, then the real-time ACUTE encoder effectively doubles the camera's raw data throughput without loss of image quality.

The key issue when image compression is done by downsampling is how to recover high frequency features.  An obvious treatment of the problem is to make downsampling scheme adaptive to local gradients and signal waveform, as suggested by Wu {\it et al.}\ \cite{wu2009low}.  This
will facilitate the recovery of details at the decoder, however, also inevitably increase the encoder complexity and hence defeat our original purpose of making the complexity of the ACUTE encoder as low as possible.  Therefore, we deliberately leave the technical challenge to the decoder of the ACUTE image compression system and meet it like in distributed source coding.  As such, all technical innovations and developments of this work are with respect to the ACUTE decoder.

The task of the ACUTE decoder appears to be one of image superresolution (SR); but unlike in common SR setting, not only the downsampling kernel is exactly known to the ACUTE decoder, more importantly the ACUTE encoder also transmits some prior information (say local gradient, as to be elaborated in Section \ref{section:Compression}) on the very latent image.  Now with the extra information and ample computation resources available to the decoder, we can strive for much higher reconstruction quality than the current state of the art methods of image superresolution.
In \cite{shu2018real}, Shu and Wu proposed an ACUTE image compression method for ultrahigh speed cameras.  They treated ACUTE decoding as a classical inverse problem and solved it via convex programming with the constraints of the prior information provided by the encoder.

In this work, encouraged by recent successes of deep learning in image restoration, we tap the power of CNN in nonlinear mapping to squeeze out extra coding gains from the ACUTE image compression paradigm.
Our main innovation is to incorporate into the CNN a hexagonal $A_2$ lattice quantizer to code the 2D gradient prior information.  Thanks to optimal space packing property and high regularity of $A_2$ lattice \cite{conway1998sphere}, it improves quantization precision without materially increasing the ACUTE encoder complexity.  More importantly, we design a soft lattice dequantization module to enforce a feasible region of all local gradients, and in this way aid the CNN superresolution module at the decoder. The two modules are end-to-end optimized in the training process.  We denote by SR-LQ$^{-1}$, the above novel CNN architecture consisting of a superresolution (SR) subnetwork and a lattice dequantization (LQ$^{-1}$) subnetwork.
To further enhance the performance of the SR-LQ$^{-1}$ network, we couple the $A_2$ lattice quantizer wih a companding mapping \cite{ogunfunmi2010principles, Sonawane2013ComparativeAB} to compensate for biases in source distributions.


Although the proposed SR-LQ$^{-1}$ CNN has a training phase that involves a computationally expensive deep learning process, at the inference stage, the decoding speed is substantially higher than the classical convex programming solver \cite{andersen2011interior}.
Unlike the pure CNN end-to-end compression methods, the proposed SR-LQ$^{-1}$ method combines deep learning and hand-crafted modeling.
To appreciate the advantages of our algorithm design, it should be noted that in foreseeable future, due to its sheer size, the CNN architecture for pure end-to-end compression \cite{nakanishi2019neural, toderici2017full, chen2021end}
cannot achieve the ultra high encoder throughput required by high frame rate cameras.

\section{Compression by Downsampling and Lattice Quantization}
\label{section:Compression}

In the proposed ACUTE image compression system, the encoder uses the basic idea presented in \cite{shu2018real}. In each $2 \times 2$ non-overlapping window of the image, the average value and diagonal gradients are computed. Let the $i$-th block in the image be
\begin{equation}
  x_i = \begin{pmatrix}
  x_{i,1} & x_{i,3}\\
  x_{i,2} & x_{i,4}
\end{pmatrix}
\end{equation}
The average value and diagonal gradients are denoted by
\begin{align}
    z_{i} &= \frac{x_{i,1}+x_{i,2}+x_{i,3}+x_{i,4}}{4} \\
    g_{i} &= x_{i,1}-x_{i,4} \\
    h_{i} &= x_{i,2}-x_{i,3}
\end{align}
To determine the appropriate number of bits for quantizing $g_i$, $h_i$, and $z_i$, we estimate the entropy of $g$, $h$, and $z$ using the DIV2K dataset. The calculated entropies for $g$, $h$, and $z$ are $5.25$, $5.25$, and $7.89$ bits, respectively. These values align with our expectations, particularly since $z$ represents the average pixel value and exhibits an approximately uniform distribution hence higher bit rate than $g$ and $h$.
Considering that most hardware systems utilize bytes (8 bits) for value storage, we assign $4$ bits (equivalent to 1 byte) to both $g$ and $h$, and allocate $8$ bits (1 byte) to $z$ to accommodate their respective quantization requirements.

\begin{figure}[H]
  \centering
  \includegraphics[width=\linewidth]{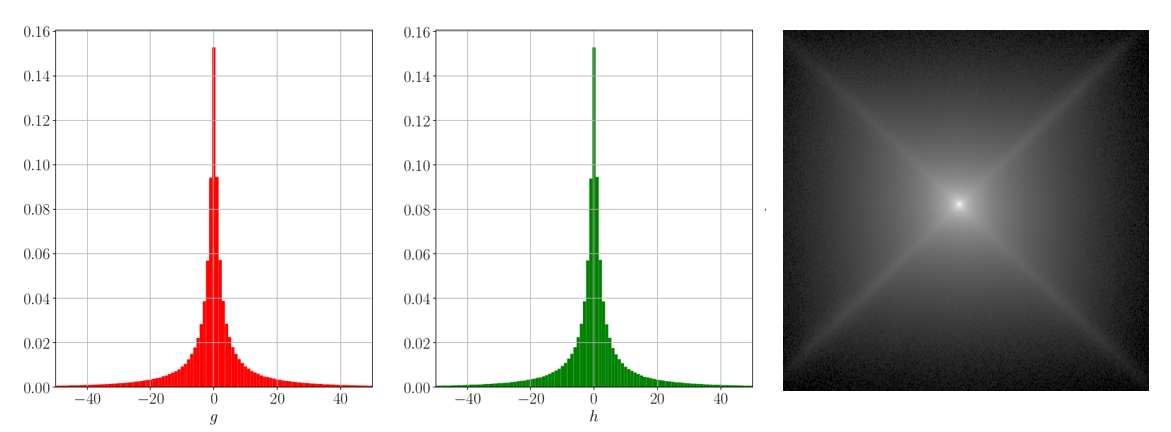}
  \caption{\textbf{Statistics of $g$ and $h$}.  Images in the DIV2K dataset \cite{agustsson2017ntire, timofte2018ntire} are sampled.
  The plots are all in log scale.}
  \label{fig:gh_hist_samples}
\end{figure}

\begin{figure}[H]
  \centering
  \begin{subfigure}[b]{0.49\linewidth}
    \includegraphics[width=\linewidth]{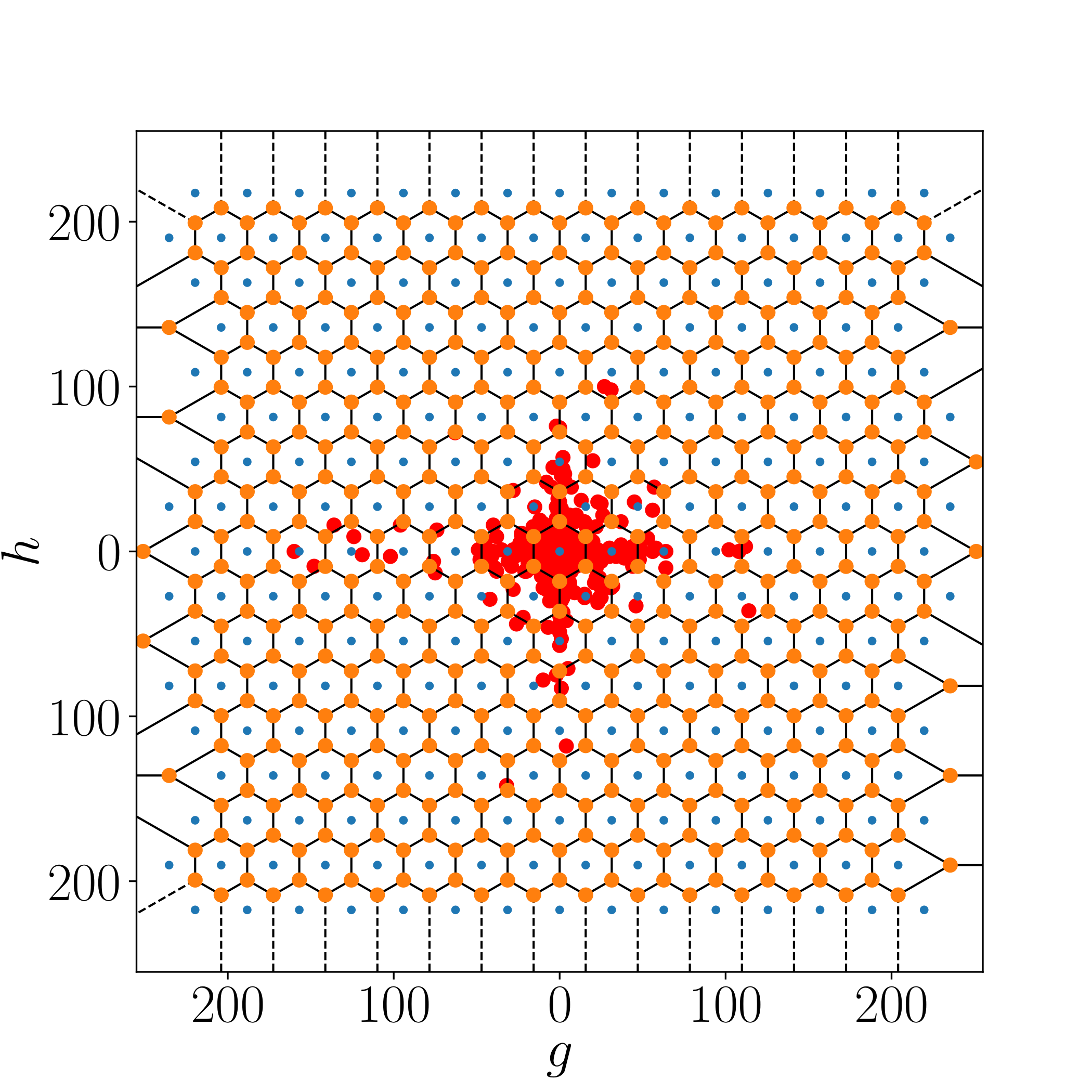}
    \caption{Data points before expansion}
  \end{subfigure}
  \begin{subfigure}[b]{0.49\linewidth}
    \includegraphics[width=\linewidth]{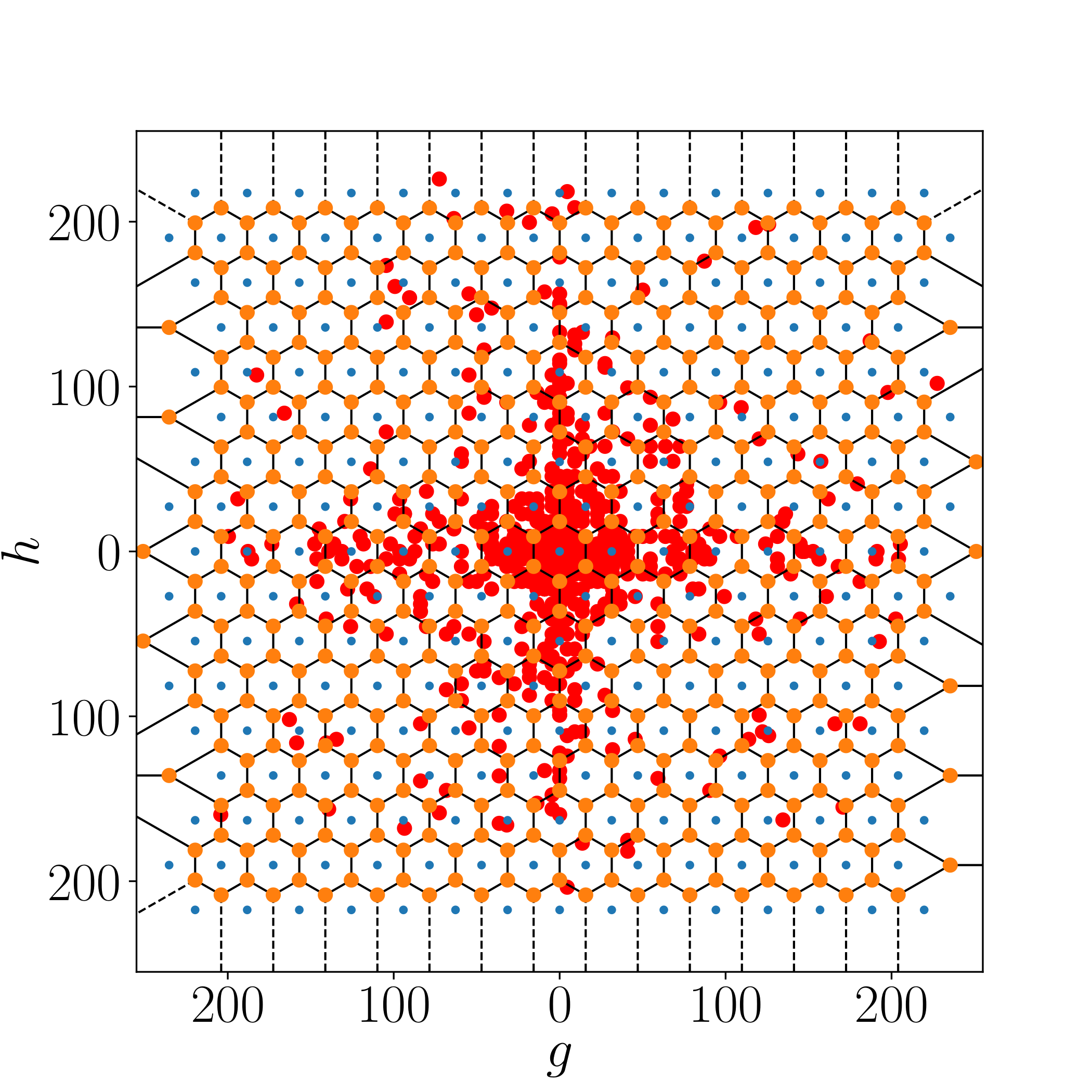}
    \caption{Data points after expansion}
  \end{subfigure}
  \caption{\textbf{Voronoi diagram of the quantizer}. Shown are the boundaries of the $A_2$ lattice quantizer and the expanded samples of $g$ and $h$.}
  \label{fig:voronoi}
\end{figure}

Figure \ref{fig:outline} presents the downsampled image $I_\downarrow$ and the corresponding gradient image $I_\nabla$.  In \cite{shu2018real} the gradient components $g_i$ and $h_i$ are each uniformly quantized into $4$ bits.  In this work we replace the old scheme by a two-dimensional hexagonal $A_2$ lattice quantizer to code the gradient vector $\nabla_i = (g_i, h_i)$ as a whole.
The regular structure of the $A_2$ vector lattice (shown in Figure \ref{fig:voronoi}) still retains low encoder complexity, while reducing quantization distortion by more efficient hexagonal space packing \cite{conway1998sphere}.  However, one problem remains, that is the mismatch between the uniform cell size of the $A_2$ lattice and the
highly skewed distribution of $\nabla_i$, as shown in Figure \ref{fig:gh_hist_samples}.


Compensating for distribution biases before quantization is one way of mitigating the coding loss of nonadaptive quantization.
A simple and yet effective technique to even out biased sample distribution is companding, such as the $mu$-law and A-law companding algorithms.  They are used to improve the dynamic range of highly biased data (e.g., speech).  For the task of ACUTE image compression, we choose the A-law algorithm to expand the gradient values near the origin, because it increases the quantization precision for values of high probability.
This not only contributes to the $\ell_2$ objective image quality but also has the perceptual benefit of making quantization distortions in smooth areas less noticeable.

\begin{figure}[H]
  \centering
  \includegraphics[width=0.99\linewidth]{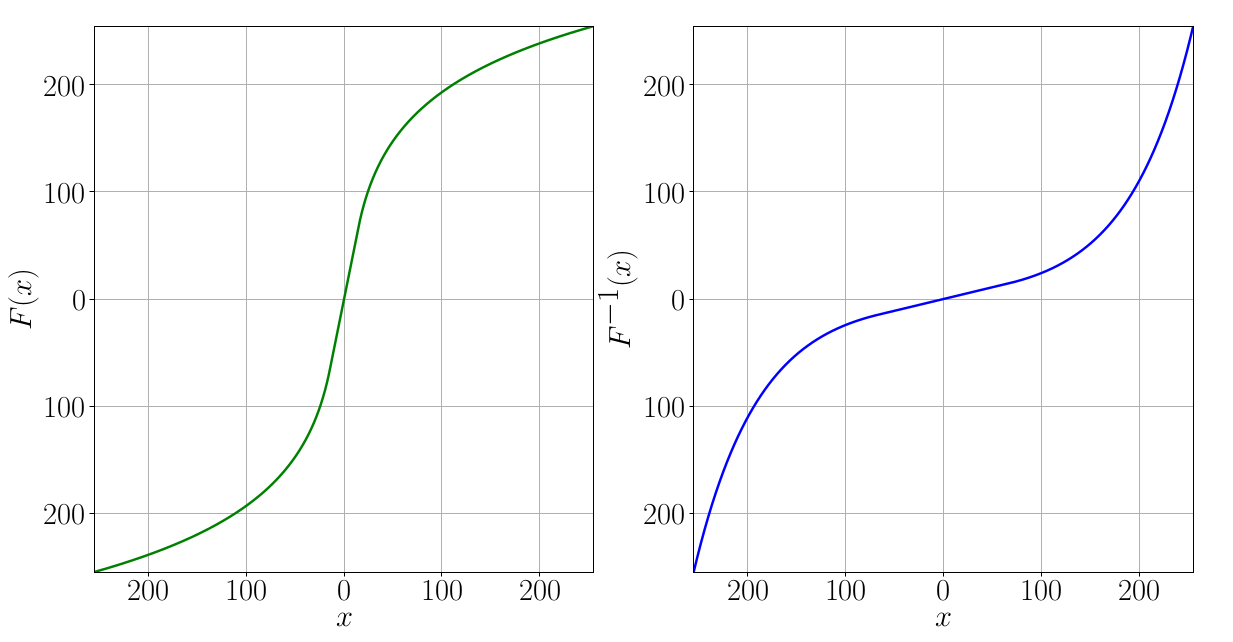}
  \caption{\textbf{A-Law companding functions for $\alpha=17.62$}. The input $x$ is unnormalized. Expansion function (left) versus  compression function (right).}
  \label{fig:companding}
\end{figure}

Specifically, we apply the following A-law companding transformation to $g$ and $h$ before quantization:
\begin{equation}
F(x)=255 \times sgn(x) \begin{dcases}
      \frac{\alpha |x|}{1+ln(\alpha)} & |x| \leq \frac{1}{\alpha}\\
      \frac{1+ln(\alpha |x|)}{1+ln(\alpha)} & \frac{1}{\alpha} \leq |x| \leq 1
   \end{dcases}
\end{equation}
where $x$ is the normalized $g$ and $h$ between $-1$ and $1$. $F(x)$ is illustrated for the unnormalized input in Figure \ref{fig:companding}.
To reverse the expanding effect of the A-law transformation done by the encoder, the decoder uses inverse function $F^{-1}$ to compress the received signal again and recover the actual values of $g$ and $h$.
To simplify notations, hereafter, the curly letters $\mathcal{G}=F(g)$ and $\mathcal{H}=F(h)$ are used to represent the expanded signals of the corresponding originals.

After the A-law mapping the next step is to quantize transformed gradient vector $\nabla_i$ in $A_2$ lattice.
To achieve 2:1 compression ratio we encode each vector $\nabla_i$ with $8$ bits, i.e., adopting a codebook of 256 lattice codewords.
Unlike other vector quantization methods, the nearest neighbor encoding of $A_2$ lattice can be carried out in constant time independent of the codebook size.
This is because the $A_2$ codewords are the centers of regularly structured hexagonal lattice cells.
The generator matrix of these $A_2$ codewords is:
\begin{equation}
    G_{A_2}=s\begin{pmatrix}
    1 & \frac{1}{2} \\
    0 & \frac{\sqrt{3}}{2}
    \end{pmatrix}
\end{equation}
In which $s$ is a hyperparameter and can scale the qunatizaion cells (hexagons) in the 2D space.
$A_2$ lattice encoding process can be simply implemented by using the uniform quantizer.
The lattice centroids are the union of two cosets linked by a global translation vector. The centroids in each coset are evenly distributed in the 2D space. Therefore, to find the corresponding centroid for an arbitrary point in the lattice quantizer, one can use two scaler quantizers to find two candidates in each coset and then pick one with the better distortion. The proposed encoding procedure for $A_2$ lattice ensures that time and space complexities are nearly as low as the uniform quantizer, while making significant performance gains as we will report and discuss in Figure \ref{fig:quantizers}.

Aiming to achieve the best possible performance under the stringent real-time throughput constraint, we jointly optimize the scaling factor $s$ and the companding parameter $\alpha$, by solving the following optimization problems in an off-line design step.
\begin{equation}
\! \! \! \begin{pmatrix}  s^* \\ \alpha^* \end{pmatrix} \! = \! \arg \! \min_{s, \alpha} \sum_{i=1}^{N} \left\| \begin{pmatrix} F^{-1}(Q(\mathcal{G}_i; s) \\ F^{-1}(Q(\mathcal{H}_i;s) \end{pmatrix} \! - \! \nabla_i \right\|_2^2
\end{equation}
Note that both $\mathcal{G}$ and $\mathcal{H}$ depends on the parameter $\alpha$.
The optimized values for $s$ and $\alpha$ are $31.38$ and $17.62$ respectively.
The Voronoi diagram of the designed quantizer are illustrated in Figure \ref{fig:voronoi}. The corresponding quantizer codebook consists of the $256$ closest centroids to the origin (in $\ell_1$ norm) produced by the generator matrix $G$.  In effect, each of $g$ and $h$ is encoded into $8$ bits by this lattice qunatizer.

\begin{figure}[H]
  \centering
  \includegraphics[width=0.8\linewidth]{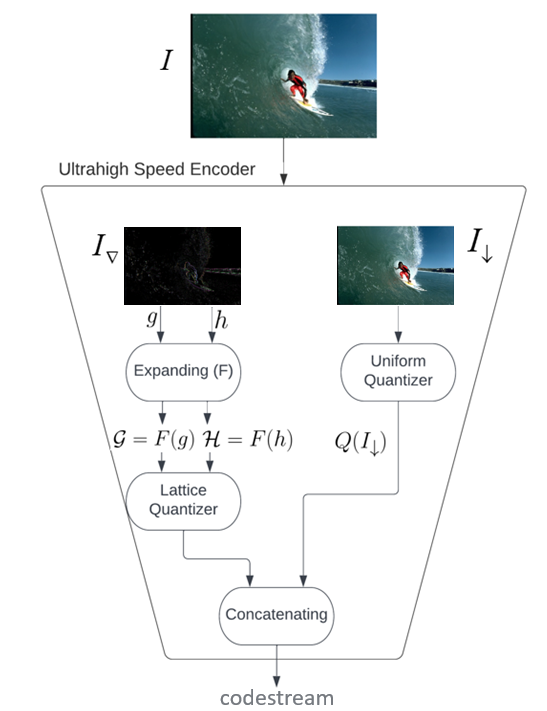}
 \caption{\textbf{Schematic diagram of the ACUTE encoder}. The input image is partitioned into $2 \times 2$ blocks. The gradient vector $(g,h)$ goes through the expanding function $F$ and then is coded by the $A_2$ lattice quantizer into 8 bits (the 2D lattice codebook has 256 codewords).
 These 8 bits are concatenated to the quantized $2 \times 2$ block average value $z$ to form the code stream.}
  \label{fig:ecndoerscheme}
\end{figure}

Finally, the encoder sends, for each $2 \times 2$ block $i$, the qunatized block average $z_i$ in $8$ bits, and the 8-bits codeword index of the corresponding gradient $\nabla_i$.
The design and algorithmic flow of the ACUTE encoder are schematically depicted in Figure \ref{fig:ecndoerscheme}.

\begin{figure*}
  \centering
  \includegraphics[width=0.9\linewidth]{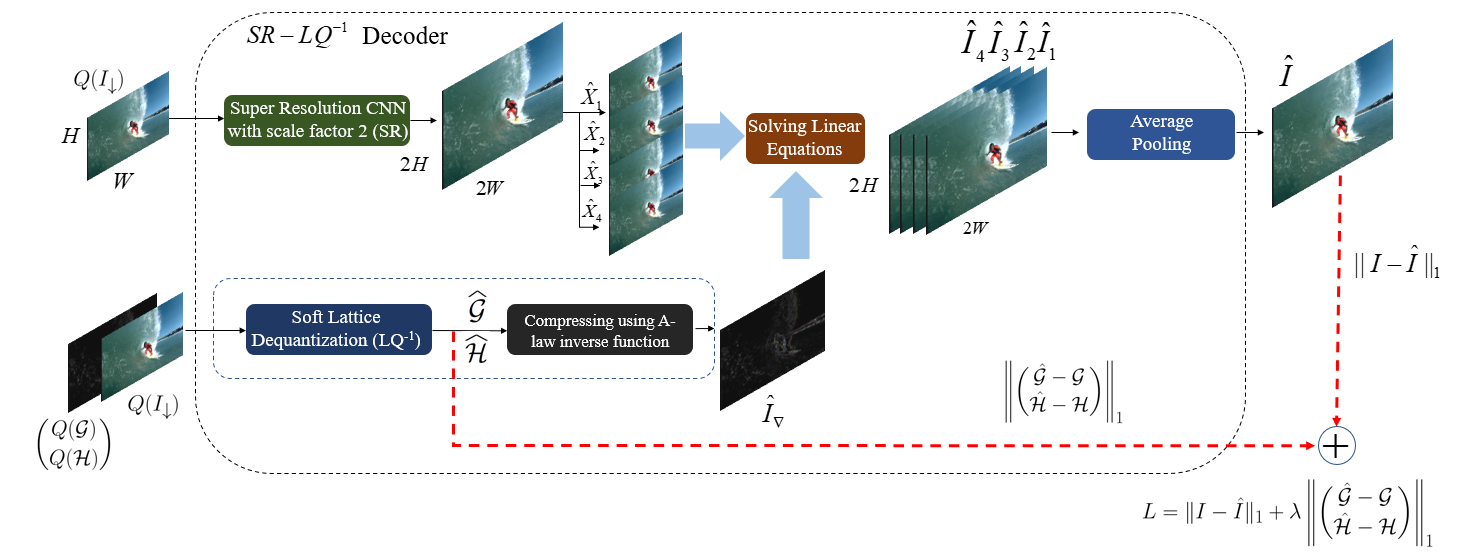}
  \caption{\textbf{Schematic diagram of the SR-LQ$^{-1}$ decoder}. The downsampled image $I_\downarrow$ is fed to a super-resolution subnetwork to make an estimate of the original resolution.
  In parallel, the lattice dequantization subnetwork LQ$^{-1}$ embarks on removing quantization errors in $Q(I_\nabla)$ to restore the gradient image $I_\nabla$.
  The results of these two subnetworks are combined to refine the final reconstruction image.}
  \label{fig:decoderschematic}
\end{figure*}

\section{SR-LQ$^{-1}$ Decoder}
Fortunately, for many applications of image/video compression, the decoder or receiver side is not limited by computational resources.  Therefore, we can exploit the power of Deep Convolutional Neural Networks to remove or alleviate compression errors and hence achieve significantly higher reconstruction fidelity than conventional decoding.
Specifically in the ACUTE compression diagram, we develop
a CNN decoder to reverse two encoding operations, down sampling and quantization.  This novel decoder is called hereafter the SR-LQ$^{-1}$ method or network for it solves the two underlying inverse problems of superresolution (SR) and lattice dequantization (LQ$^{-1}$) and optimize the solutions jointly.  Unlike conventional decoders, the SR-LQ$^{-1}$ method can be
a highly complex nonlinear mapping, if so required to recover the original image best possible.  Also, it should be noted that the
LQ$^{-1}$ lattice dequantization is a more sophisticated soft decoding operation than the conventional hard decoding of simply choosing the center of lattice cell.

In designing the SR-LQ$^{-1}$ decoder, the challenge is how to maximally profit from the prior knowledge on the local average $z_i$ and gradient $\nabla_i$ of each pixel $i$.  These priors tie the four samples $x_1$, $x_2$, $x_3$ and $x_4$ by three linear equations.  Such strong constraints cannot be exploited by existing super-resolution CNN architectures when upsampling the local average value of the original image; even the best of SR CNNs do not respect, in general, the priors of $z_i$ and $\nabla_i$.
In the next subsections, we present the SR-LQ$^{-1}$ CNN decoder in detail and justify our architecture design and optimization strategy.
As schematically illustrated in Figure \ref{fig:decoderschematic},
the SR-LQ$^{-1}$ CNN decoder consists of two subnetworks: (i) a super-resolution CNN (SR) and (ii) a Soft Lattice Dequantization CNN (LQ$^{-1}$).

\subsection{Super-resolution Module}
In the SR-LQ$^{-1}$ method, any SR CNN can play the role of upsampling the $2 \times 2$ block average $z$ to $x_1$, $x_2$, $x_3$, and $x_4$.  We choose the EDSR CNN, one of the best known SR architectures, in the place of SR module in our SR-LQ$^{-1}$ design.
As well known, SR results, regardless produced by traditional or CNN methods, tend to be somewhat too smooth and less capable of recovering sharp details, representing a low-pass approximations of the original HR image.  Precisely for overcoming this difficulty and recovering high-frequency features, the ACUTE encoder sends the quantized gradient image $\hat{I}_\nabla$
to aid the decoder.



The SR-LQ$^{-1}$ decoder does not use the quantized gradient image
$Q(I_\nabla)$ directly when refining high frequency details.  Instead, it tries to reduce quantization errors and use an improved estimate $\hat{I}_\nabla$ of the gradient image $I_\nabla$.
To recover $I_\nabla$, we design a soft lattice dequantization LQ$^{-1}$ subnetwork that is jointly end-to-end optimized with the SR subnetwork.
This is the main technical innovation of SR-LQ$^{-1}$ to be detailed in the following subsection.



\subsection{Soft Lattice Dequantization}

\begin{figure}[H]
  \centering
  \includegraphics[width=\linewidth]{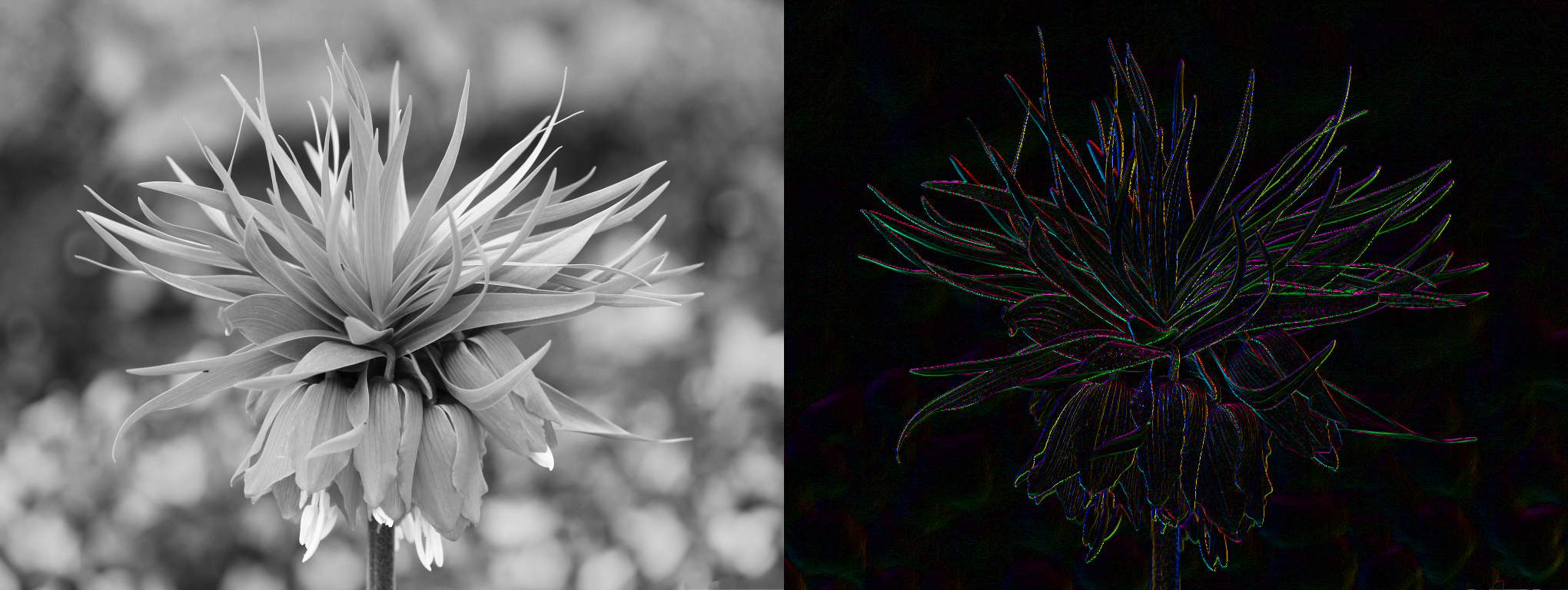}
  \caption{\textbf{Structural correlation between $I_\downarrow$ and $I_\nabla$}. The gradient image has both angle and magnitude which is color coded by value and hue in HSV space respectively. The presence of specific structures and shapes in $I_\nabla$ exhibits the local correlation in the gradient image. Moreover, the similarity between $I_\downarrow$ and $I_\nabla$ shows the cross-correlation between these two images. }
  \label{fig:corr}
\end{figure}

One can easily observe from Figure \ref{fig:corr} that the downsampled image $I_\downarrow$ and the corresponding gradient image $I_\nabla$ have structural cross correlations, and also local correlations exist within $I_\nabla$ itself.  Accordingly, we set out to restore the gradient image $I_\nabla$ from its lattice quantized version $Q(I_\nabla)$ via supervised deep learning, and use the restored gradient image $\hat{I}_\nabla$ to facilitate the final CNN reconstruction of the latent image $I$.

\begin{figure*}
  \centering
  \includegraphics[width=0.8\linewidth]{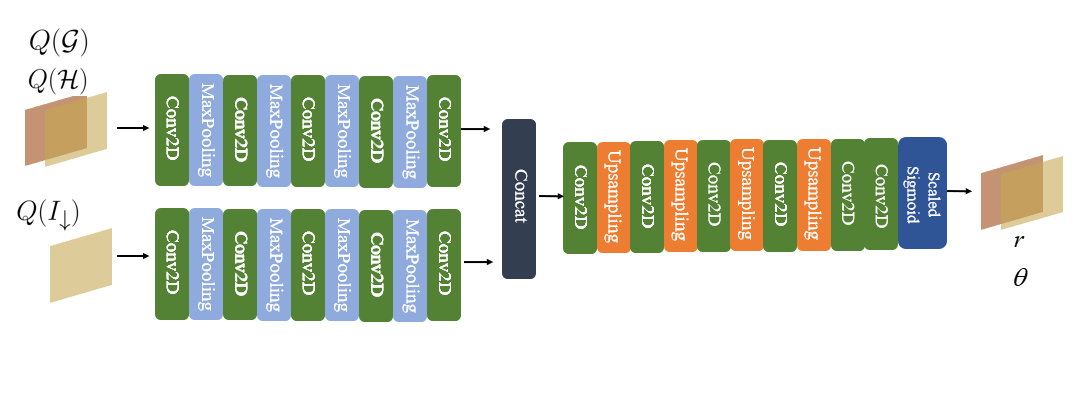}
  \caption{\textbf{The architecture of soft lattice dequantization LQ$^{-1}$}. LQ$^{-1}$ extracts features from quantized downsampled image $Q(I_\downarrow)$ and lattice quantized gradient image $Q(I_\nabla)$, separately.  These two sets of features are then fused by convolution and upsampling layers to estimate the original gradient vector $(r,\theta)$ in polar coordinates.
  }
  \label{fig:qe}
\end{figure*}

\begin{figure}
  \centering
  \includegraphics[width=\linewidth]{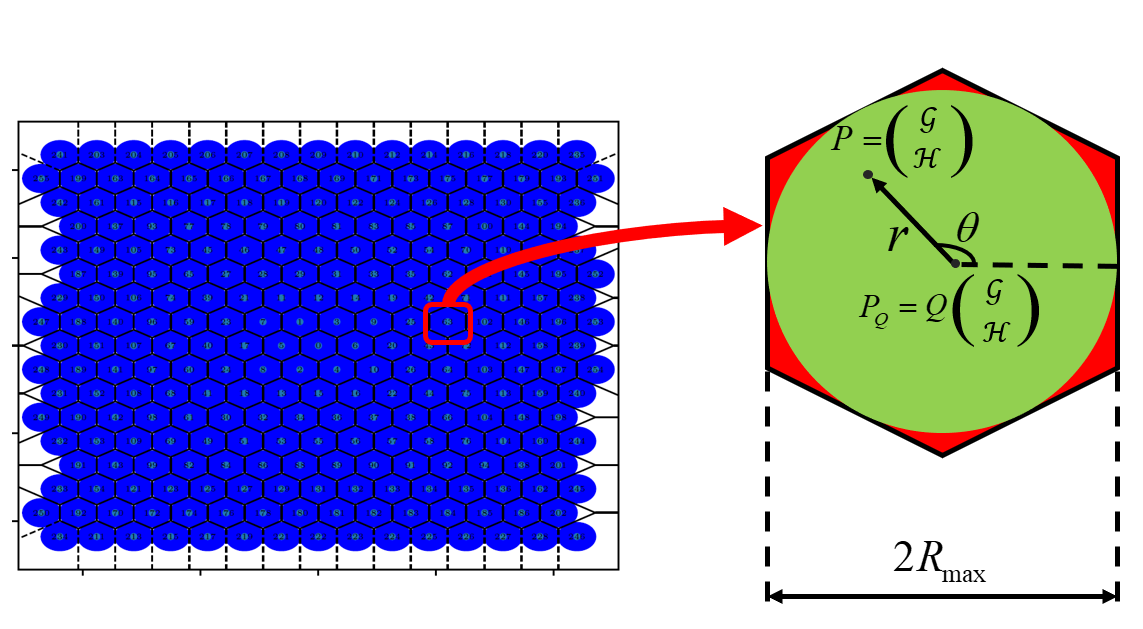}
  \caption{\textbf{Left: lattice quantizer cells; Right: the inscribed circle for one of the cells}. In the right image, $P$ is the actual expanded gradient vector $(\mathcal{G},\mathcal{H})$ 
  in terms of $r$ and $\theta$; the cell center $P_Q$ is the quantized $P$ $(\hat{\mathcal{G}},\hat{\mathcal{H}})$. }
  \label{fig:cells}
\end{figure}

Arguably, even more beneficial is the lattice structure of the $A_2$ quantizer, which effectively confines the solution space for gradient vector $\nabla_i$ at each pixel $i$.
As depicted in Figure \ref{fig:cells} (left), after the expanding operation, the actual vector $(\mathcal{G}_i,\mathcal{H}_i)$ is bounded by the hexagon quantizer cell.  For algorithm amenability, we approximate this hexagonal domain by its inscribing circle with a negligible error near the vertices.
In the training of the soft lattice dequantization LQ$^{-1}$ CNN, we impose the above circular feasibility region.
For easy implementation of the circular constraint we switch to
polar coordinates of two parameters, radius $r$ and angle $\theta$, where the origin of the coordinate system is the center of the inscribed circle of each quantizer cell.
This approximation can introduce small errors for points close to the hexagon vertices, and for the points laying outside of the maximum and minimum allowable values for the quantization (boundaries). However, the probabilities of these occurrences are negligible.

The proposed LQ$^{-1}$ network module estimates the true gradient vector after expanding operation $F$
in polar coordinates $(r,\theta)$, namely,
\begin{align}
    \reallywidehat{\mathcal{G}} &= Q(\mathcal{G})+rcos(\theta) \\
    \reallywidehat{\mathcal{H}} &= Q(\mathcal{H})+rsin(\theta)
\end{align}
and back to the original scale
by applying $F^{-1}$:
\begin{align}
    \hat{g} &= F^{-1}(\reallywidehat{\mathcal{G}}) \\
    \hat{h} &= F^{-1}(\reallywidehat{\mathcal{H}})
\end{align}
The valid range for $r$ is $[0,R_{max}]$, where $R_{max}$ is the radius of the inscribed circle of each cell of the lattice quantizer (our optimal design of Eq(7) finds $R_{max}=15.69$).
The $\theta$ value is in the range $[0,2\pi]$.
The lower and upper bounds on $r$ can be conveniently implemented in the $LQ^{-1}$ CNN architecture by using the scaled sigmoid as the activation function in the subnetwork's last layer.
The architecture of the $LQ^{-1}$ module is illustrated in Figure \ref{fig:qe}.

\subsection{Solving Linear Equations}

For the $i$-th $2 \times 2$ pixel block, each of the values $z_i$, $g_i$ and $h_i$ defines a linear equation in terms of the four pixel values $x_{i,1}$, $x_{i,2}$, $x_{i,3}$ and $x_{i,4}$.  We have three equations for four unknowns.  If the value of one pixel in the block is fixed, the other pixels can be determined uniquely by solving the linear equations for that block.  Therefore, there are four different solutions depending on which pixel value in block $i$ is known:
\begin{equation}
    \hat{I}_1 = \begin{pmatrix}
    x_{i,1} \\
    2z_{i}-x_{i,1} + \frac{g_i+h_i}{2} \\
    2z_{i}-x_{i,1} + \frac{g_i-h_i}{2} \\
    x_{i,1}-g_i
    \end{pmatrix} \; \hat{I}_2 = \begin{pmatrix}
    2z_{i}-x_{i,2} + \frac{h_i+g_i}{2} \\
    x_{i,2} \\
    x_{i,2} - h_i \\
    2z_{i}-x_{i,2} + \frac{h_i-g_i}{2}
    \end{pmatrix}
\end{equation}
\begin{equation}
    \hat{I}_3 = \begin{pmatrix}
    2z_{i}-x_{i,3} + \frac{g_i-h_i}{2} \\
   x_{i,3} + h_i \\
   x_{i,3} \\
    2z_{i}-x_{i,3} - \frac{g_i+h_i}{2}
    \end{pmatrix} \; \hat{I}_4 = \begin{pmatrix}
    x_{i,4} + g_i \\
  2z_{i}-x_{i,4} + \frac{h_i-g_i}{2} \\
  2z_{i}-x_{i,4} - \frac{h_i+g_i}{2} \\
   x_{i,4}
    \end{pmatrix}
\end{equation}

$\hat{I}_j$, $j=1,2,3,4$, is an estimate of
block $i$ by setting the value of $x_{i,j}$ to the output value of
the upsampling network at location $j$ of block $i$ and using the quantized block average $z_i$ and the estimated gradient vector $(\hat{g}_i,\hat{h}_i)$.
We average the above four estimated images to obtain the final soft  decoded image.  This fusion strategy is motivated by our observation
that all the operations in the proposed soft decoder work flow are differentiable, thus we can train the decoder network end-to-end with an appropriate loss function (refer to Figure \ref{fig:decoderschematic}).  
In the training stage, we use both the original image $I$ and the gradient image $(g,h)$ to supervise the learning of the SR-LQ$^{-1}$ model.
Specifically, the loss function per image in terms of $\hat{I}_j$ and the restored expanded gradients
$\reallywidehat{\mathcal{G}}$ and $\reallywidehat{\mathcal{H}}$
is
\begin{align}
    \nonumber L(I, \hat{I})\! = \! ||I-\frac{1}{4}\sum_{j=1}^{4}\hat{I}_j||_1 + \lambda \left\| \! \begin{pmatrix} \reallywidehat{\mathcal{G}} - \mathcal{G} \\ \reallywidehat{\mathcal{H}} - \mathcal{H}  \end{pmatrix} \! \right\| _1
\end{align}
where $I$ is the ground truth, $\hat{I}$ is the output of the SR-LQ$^{-1}$ model and $\lambda$ determines the importance of the recovery of true gradients.

\section{Experiments and Evaluations}

This section presents empirical evidences for the competitive rate-distortion performance of the proposed ACUTE CNN soft decoding system despite the extremely simplistic encoder.
We compare ACUTE with three competing image compression methods of two categories.  In the first category, the compression and decompression are simply downsampling and upsampling operations, like ACUTE.
Two different methods of this category are included in the comparison group.  The first one is to replace the SR-LQ$^{-1}$ decoder by a deep learning-based upsampler, the EDSR network \cite{lim2017enhanced}, with the upsampling factor of $2$.  EDSR is among the best performing image super-resolution CNN models.  The second method shares the same encoder as ACUTE, but it adopts a convex programming approach to solve the inverse problem of soft decoding \cite{shu2018real}, instead of deep learning.  The second category includes traditional image compression methods without machine learning in either the encoder or the decoder. JPEG is selected to represent this category because it is most widely used and it offers both good rate-distortion performance and high speed.

Various datasets are used to validate the generalization capability of the learnt SR-LQ$^{-1}$ soft decoder, which is trained by the DIV2K dataset, over different types of images.  They are the DIV2K validation set (excluded from the training set), BSD100 \cite{martin2001adatabseof}, Set 14 \cite{bevilacqua2012low}, Set5 \cite{zeyde2012on}, CLIC \cite{toderici2020workshop} and KODAK \cite{kodak}.
Two well-known image quality metrics, PSNR and SSIM, are used to evaluate the quality of the decoded images.

\begin{figure*}
  	\begin{tabular}{ccccc}
  		 Original Image & EDSR &  \cite{shu2018real}  & ACUTE (Ours) & JPG (QF=90)\\
  		 \includegraphics[trim=0pt 0pt 0pt 0pt, clip, width=2in]{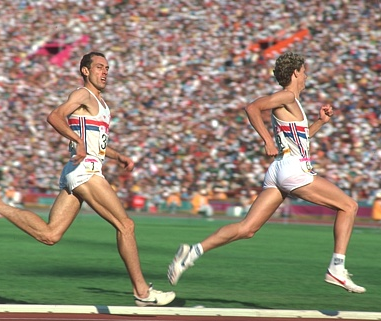} &
  		 \raisebox{1cm}{\includegraphics[trim=0pt 0pt 0pt 0pt, clip, width=1in]{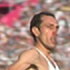}}
     	 &
 \raisebox{1cm}{\includegraphics[trim=0pt 0pt 0pt 0pt, clip, width=1in]{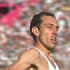}} &
     	 \raisebox{1cm}{\includegraphics[trim=0pt 0pt 0pt 0pt, clip, width=1in]{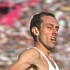}} &
     	 \raisebox{1cm}{\includegraphics[trim=0pt 0pt 0pt 0pt, clip, width=1in]{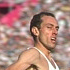}} \\
     	  &
     	  \textcolor{Blue}{27.37} /  \textcolor{red}{0.94} &   \textcolor{Blue}{31.56} /  \textcolor{red}{0.97} & \textcolor{Blue}{37.56} /  \textcolor{red}{0.99} &
     	  \textcolor{Blue}{38.67} /  \textcolor{red}{0.99}
     	  \\
  	\end{tabular} \\
  	
  	\begin{tabular}{ccccc}
  		 \includegraphics[trim=0pt 0pt 0pt 0pt, clip, width=2in]{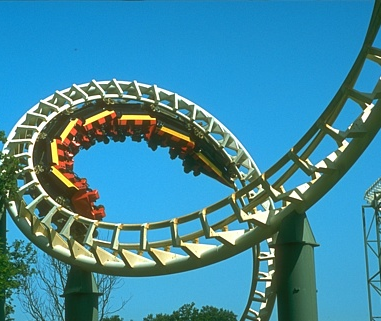} &
  		 \raisebox{1cm}{\includegraphics[trim=0pt 0pt 0pt 0pt, clip, width=1in]{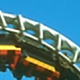}}
     	 &
 \raisebox{1cm}{\includegraphics[trim=0pt 0pt 0pt 0pt, clip, width=1in]{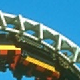}} &
     	 \raisebox{1cm}{\includegraphics[trim=0pt 0pt 0pt 0pt, clip, width=1in]{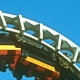}} &
     	 \raisebox{1cm}{\includegraphics[trim=0pt 0pt 0pt 0pt, clip, width=1in]{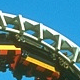}} \\
     	  &
     	  \textcolor{Blue}{26.60} /  \textcolor{red}{0.92} &   \textcolor{Blue}{28.32} /  \textcolor{red}{0.96} & \textcolor{Blue}{35.73} /  \textcolor{red}{0.99} &
     	  \textcolor{Blue}{38.25} /  \textcolor{red}{0.98}
     	  \\
  	\end{tabular}
  	
  	\begin{tabular}{ccccc}
  		 \includegraphics[trim=0pt 0pt 0pt 0pt, clip, width=2in]{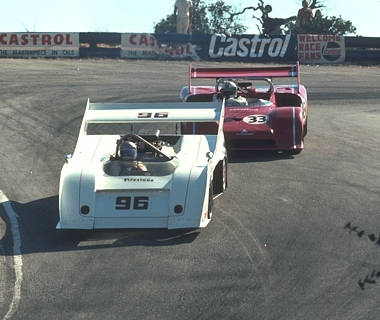} &
  		 \raisebox{1cm}{\includegraphics[trim=0pt 0pt 0pt 0pt, clip, width=1in]{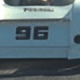}}
     	 &
 \raisebox{1cm}{\includegraphics[trim=0pt 0pt 0pt 0pt, clip, width=1in]{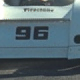}} &
     	 \raisebox{1cm}{\includegraphics[trim=0pt 0pt 0pt 0pt, clip, width=1in]{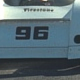}} &
     	 \raisebox{1cm}{\includegraphics[trim=0pt 0pt 0pt 0pt, clip, width=1in]{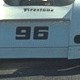}} \\
     	  &
     	  \textcolor{Blue}{31.77} /  \textcolor{red}{0.93} &   \textcolor{Blue}{34.65} /  \textcolor{red}{0.98} & \textcolor{Blue}{41.20} /  \textcolor{red}{0.99} &
     	  \textcolor{Blue}{40.48} /  \textcolor{red}{0.98}
     	  \\
  	\end{tabular}

  	\begin{tabular}{ccccc}
  		 \includegraphics[trim=0pt 0pt 0pt 0pt, clip, width=2in]{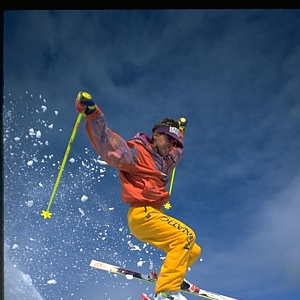} &
  		 \raisebox{1cm}{\includegraphics[trim=0pt 0pt 0pt 0pt, clip, width=1in]{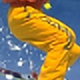}}
     	 &
 \raisebox{1cm}{\includegraphics[trim=0pt 0pt 0pt 0pt, clip, width=1in]{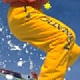}} &
     	 \raisebox{1cm}{\includegraphics[trim=0pt 0pt 0pt 0pt, clip, width=1in]{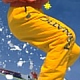}} &
     	 \raisebox{1cm}{\includegraphics[trim=0pt 0pt 0pt 0pt, clip, width=1in]{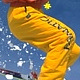}} \\
     	  &
     	  \textcolor{Blue}{24.01} /  \textcolor{red}{0.83} &   \textcolor{Blue}{28.21} /  \textcolor{red}{0.94} & \textcolor{Blue}{33.06} /  \textcolor{red}{0.98} &
     	  \textcolor{Blue}{37.13} /  \textcolor{red}{0.98}
     	  \\
  	\end{tabular}
  	
	\caption{\textbf{Visual comparison of different compression methods for scenes of very fast motions (\textcolor{Blue}{PSNR}/\textcolor{red}{SSIM}).}}
	\label{fig:res2}
\end{figure*}

\setlength{\tabcolsep}{3pt}
\begin{table}[H]
\centering
\caption{\footnotesize \label{tab:results} \textbf{Comparison of different compression methods in quantitative image quality for common test image sets.}}
\label{tab:performance}
\begin{tabular}{c|c|c|ccc|c}
 & EDSR & \cite{shu2018real} &  & JPEG &  & ACUTE \\
 & \cite{lim2017enhanced} & & QF=85 & QF=90 & QF=92 & \\
\hline
\textcolor{red}{\textbf{DIV2K}} & & & & & &  \\
PSNR & 33.40 & 36.59 & 39.26 & 41.27 & 42.42 & 41.82 \\
SSIM & 0.93 & 0.97 & 0.97 & 0.98 & 0.98 & 0.99  \\
Ratio & 4 & 2 & 3.19 & 2.5 & 2.26 & 2 \\
[0.3cm]
\hline
\textcolor{red}{\textbf{BSD100}}& & & & & &  \\
PSNR & 29.96 & 33.96 & 37.33 & 38.81 & 42.04 & 39.08 \\
SSIM & 0.89 & 0.96 & 0.97 & 0.98 & 0.99 & 0.99  \\
Ratio & 4 & 2 & 2.76 & 2.05 & 1.85 & 2  \\
[0.3cm]
\hline
\textcolor{red}{\textbf{Set14}} & & & & & &  \\
PSNR & 31.32 & 34.22 &  38.54 & 41.25 & 42.66 & 39.68 \\
SSIM & 0.90 & 0.96 & 0.96 & 0.98 & 0.98 & 0.99 \\
Ratio & 4 & 2 & 2.73 & 2.17 & 1.98 & 2 \\
[0.3cm]
\hline
\textcolor{red}{\textbf{Set5}} & & & & & & \\
PSNR & 35.66 & 37.55 & 40.49 & 42.38 & 44.0 & 42.97 \\
SSIM & 0.96 & 0.98 & 0.98 & 0.98 & 0.98 & 0.99 \\
Ratio & 4 & 2 & 2.92 & 2.37 & 2.19 & 2  \\
[0.3cm]
\hline
\textcolor{red}{\textbf{CLIC (test)}} & & & & & & \\
PSNR & 35.61 & 38.13 & 40.61 & 42.35 & 43.33 & 42.87 \\
SSIM & 0.94 & 0.97 & 0.97 & 0.98 & 0.98 & 0.99 \\
Ratio & 4 & 2 & 3.77 & 2.87 & 2.57 & 2  \\
[0.3cm]
\hline
\textcolor{red}{\textbf{CLIC (val)}} & & & & & & \\
PSNR & 35.06 & 38.50 & 40.13 & 41.91 & 42.91 & 43.16 \\
SSIM & 0.93 & 0.97 & 0.97 & 0.98 & 0.98 & 0.99 \\
Ratio 4 & 2 & 2 & 3.54 & 2.72 & 2.44 & 2  \\
[0.3cm]
\hline
\textcolor{red}{\textbf{KODAK}} & & & & & & \\
PSNR & 31.67 & 35.40 & 38.18 & 40.23 & 41.42 & 40.83 \\
SSIM & 0.90 & 0.97 & 0.96 & 0.97 & 0.98 & 0.99 \\
Ratio & 4 & 2 & 2.97 & 2.33 & 2.11 & 2  \\
[0.3cm]
\hline
\textbf{Average} & & & & & & \\
PSNR & 33.24 & 36.33 & 39.22 & 41.17 & 42.68 & 41.48  \\
SSIM & 0.92 & 0.96 & 0.96 & 0.97 & 0.98 & 0.99 \\
Ratio & 4 & 2 & 3.12 & 2.43 & 2.2 & 2 \\
\end{tabular}
\end{table}

\subsection{Experiment Setting}
The upsampler module in SR-LQ$^{-1}$ has the same architecture as EDSR with upscaling factor 2.  Randomly selected $256 \times 256$ patches are used to train the SR-LQ$^{-1}$ network; the training is carried out by the Yogi optimizer \cite{zaheer2018adaptive}, with an initial learning rate of $10^{-4}$.
The training process takes $50000$ iterations and sets the minibatch size to $8$.

At the inference time, we use the self-ensembling technique similar to the proposed method in \cite{lim2017enhanced} to increase the performance of ACUTE.  Specifically, each test image goes through multiple geometric transforms (rotation by $0, 90, 180, 270$ degrees, flipping vertically and horizontally) to have different versions.
Each of these transformed versions is fed to the SR-LQ$^{-1}$ network and restored.  Then the inverse geometrical transform is applied to bring each restored version back to the original geometry.  The median over all of these restored results is taken as the final output.

When compressing the images with the JPEG algorithm, we run for three different quality factors, 85, 90, and 92. The quality factor of 85 is typically used as the default value in practice. The quality factors of 90 and 92 roughly give the average compression ratio of 2:1 for the natural images.  It should be stressed that JPEG has much higher encoder complexity than the encoder of ACUTE.  The former is not suited for extremely high throughput imaging devices, such as very high frame rate video cameras, while the latter is.

\subsection{Comparing encoding time: JPEG vs. ACUTE}
In this section, we compare the required times to encode and $8 \times 8$ block of the image by JPEG and ACUTE.
Assume the readout process takes $t_r$ to buffer a line of an image. In JPEG, the major computation is in performing DCT. Computing DCT on $8 \times 8$ block needs $64$ multiplications and $224$ additions, if we assume the time for each floting point operation is $t_a$, then the time for computing DCT is $288t_a$. Finally, JPEG needs to use the entropy coder to encode the values. The entropy encoder involves counting the occurrences of different DCT coefficients and performing arithmetic operations to encode them. Assuming the required time for the entropy encoder is $t_h$. The total time to encode an $8 \times 8$ block by JPEG is $8t_r+288t_a+t_h$. In ACUTE encoder, we only need to buffer $2$ lines of the image to start encoding process. Therefore by parallelizing, the encoding time is the same as the encoding a $2 \times 8$ block of the image. $g$, $h$ requires $2$ additions and $z$ requires $4$ additions and $1$ multiplications for a $2 \times 2$ block. Therefore for encoding the $2 \times 8$ block, it takes $28t_a+2t_r$ which is the same as encoding time for $8 \times 8$ block. The ratio of the encoding time of ACUTE to JPEG is:
\begin{equation}
    \mathrm{Gain} = \frac{T_\mathrm{JPG}}{T_\mathrm{ACUTE}} = \frac{8t_r+288t_a+t_h}{28t_a+2t_r} \geq \frac{8t_r+288t_a}{28t_a+2t_r}
\end{equation}
In recent high speed cameras, the readout process for an entire frame is in the range of Million frame per second, which means that for HD resolution, the readout time per line of the image is roughly 1 nanoseconds, which is roughly the same as the computations required for arithmetic operations in the processors. We have plotted the lower bound of the gain for different ratio in Figure \ref{fig:gain}.

\begin{figure}
  \centering
  \includegraphics[width=0.7\linewidth]{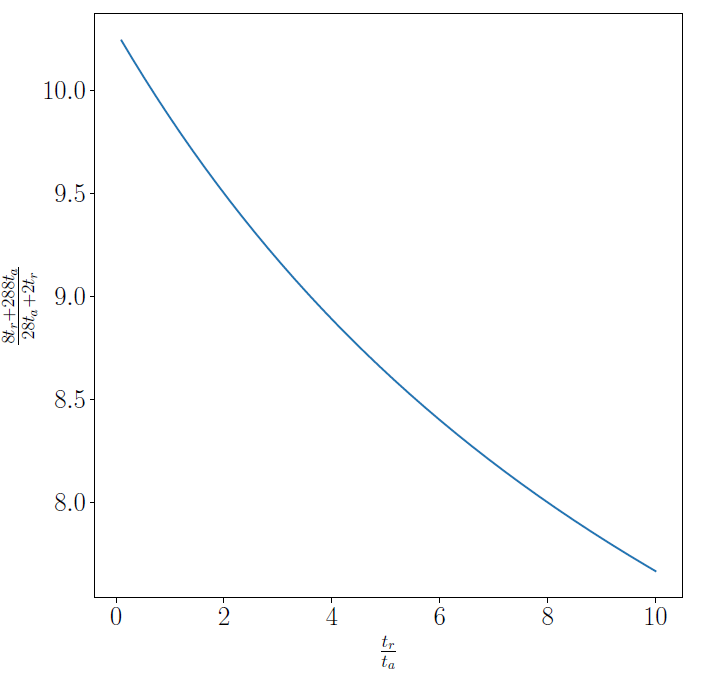}
  \caption{Gain vs $\frac{t_r}{t_a}$ For typical ultra high frame rate cameras the read out time for one line of the image is roughly equal to the floating point operation time  ($\frac{t_r}{t_a} \approx 1$), in such case ACUTE encoding time is more than $9$ times faster than JPEG encoding time.}
\end{figure}

\subsection{Results}
The performance results of tested compression methods in quantitative quality metrics of PSNR and SSIM are tabulated in Table \ref{tab:results}.
As exhibited, our method outperforms EDSR, \cite{shu2018real} and JPEG (QF=85) for all the datasets, which proves the generalization capability of ACUTE.  ACUTE even surpasses JPEG of a high-quality factor (QF=90) for all of the datasets except Set14.
For QF=92, as can be seen, the JPEG starts to beat ACUTE in reconstruction quality, while yielding compression ratio around 2:1.
However, keep in mind that the JPEG algorithm is too complex and too slow to run in ultra-high speed camera data processing pipeline.

A comparison between the results of EDSR and ACUTE indicates that sending gradient information can significantly improve the quality of the decoded images.

Figure \ref{fig:res2} compares different compression methods in visual quality.  We choose scenes of high speed movements considering that the ultra-high throughput ACUTE encoder is designed for sustained shooting of high frame rate videos.  As illustrated, our method can preserve important details for fast-moving objects (e.g., edges and fine textures) better than other methods, with a greater degree of sharpness and clarity.

\subsection{Ablation Study}
In this section, we conduct various experiments to verify the role and evaluate the effectiveness of the individual components of the ACUTE system.

\subsubsection{Soft Lattice Dequantization Module (LQ$^{-1}$)}

\begin{figure*}[!ht]
  \captionsetup[subfigure]{labelformat=empty}
  \centering
    \begin{subfigure}[b]{0.235\linewidth}
    \caption{$I$}
    \includegraphics[width=\linewidth]{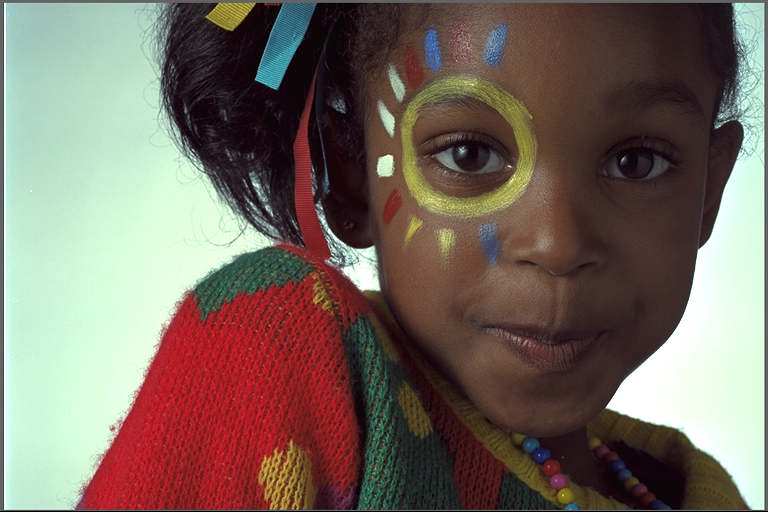}
     \caption{}
  \end{subfigure}
  \begin{subfigure}[b]{0.24\linewidth}
    \caption{$\begin{pmatrix}  Q(g) \\ Q(h) \end{pmatrix}$}
    \includegraphics[width=\linewidth]{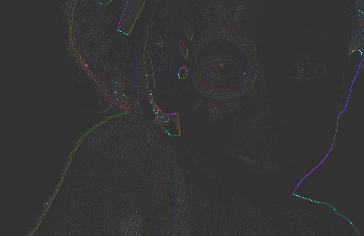}
    \caption{\textit{MAE}=4.98}
  \end{subfigure}
  \begin{subfigure}[b]{0.24\linewidth}
    \caption{$\begin{pmatrix}  \hat{g} \\ \hat{h} \end{pmatrix}$}
    \includegraphics[width=\linewidth]{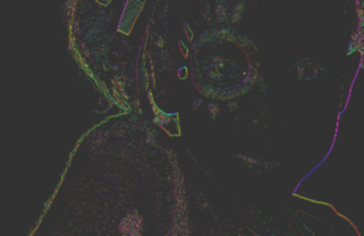}
    \caption{\textit{MAE}=1.35}
  \end{subfigure}
    \begin{subfigure}[b]{0.24\linewidth}
    \caption{$\begin{pmatrix}  g \\ h \end{pmatrix}$}
    \includegraphics[width=\linewidth]{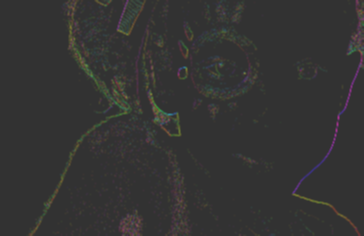}
    \caption{}
  \end{subfigure}
    \begin{subfigure}[b]{0.24\linewidth}
    \includegraphics[width=\linewidth]{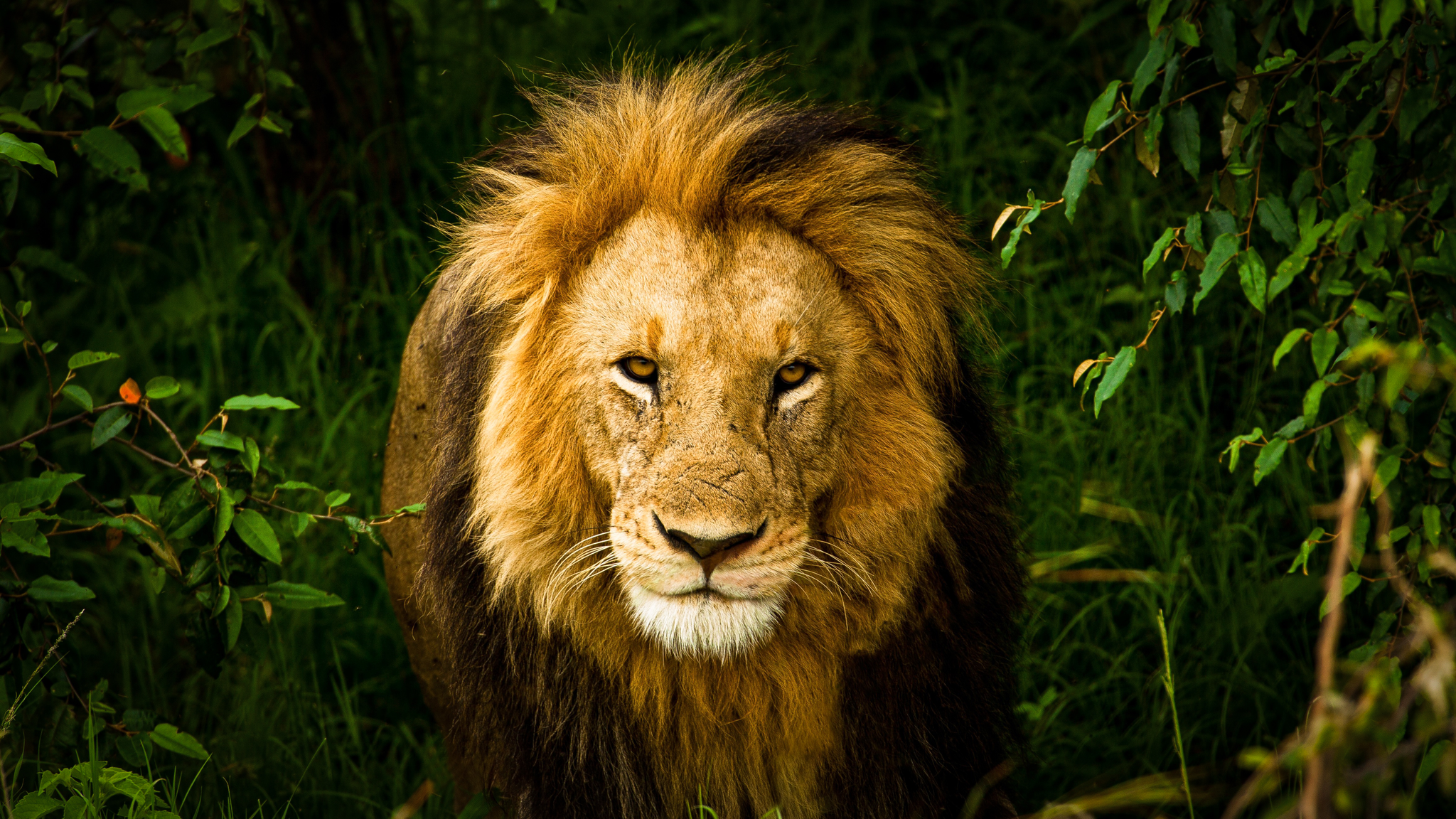}
     \caption{}
  \end{subfigure}
    \begin{subfigure}[b]{0.24\linewidth}
    \includegraphics[width=\linewidth]{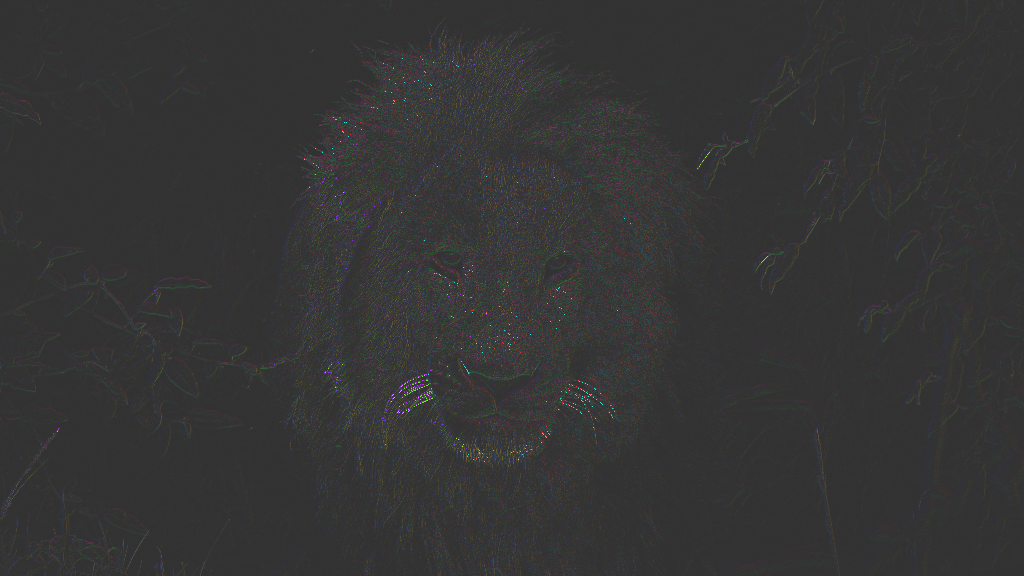}
    \caption{\textit{MAE}=6.17}
  \end{subfigure}
  \begin{subfigure}[b]{0.24\linewidth}
    \includegraphics[width=\linewidth]{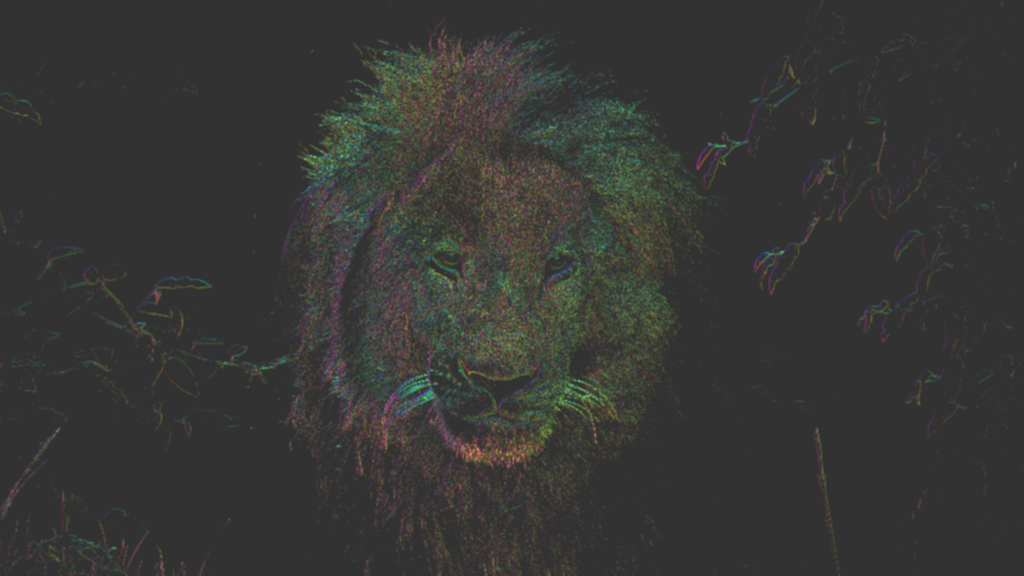}
    \caption{\textit{MAE}=1.62}
  \end{subfigure}
    \begin{subfigure}[b]{0.24\linewidth}
    \includegraphics[width=\linewidth]{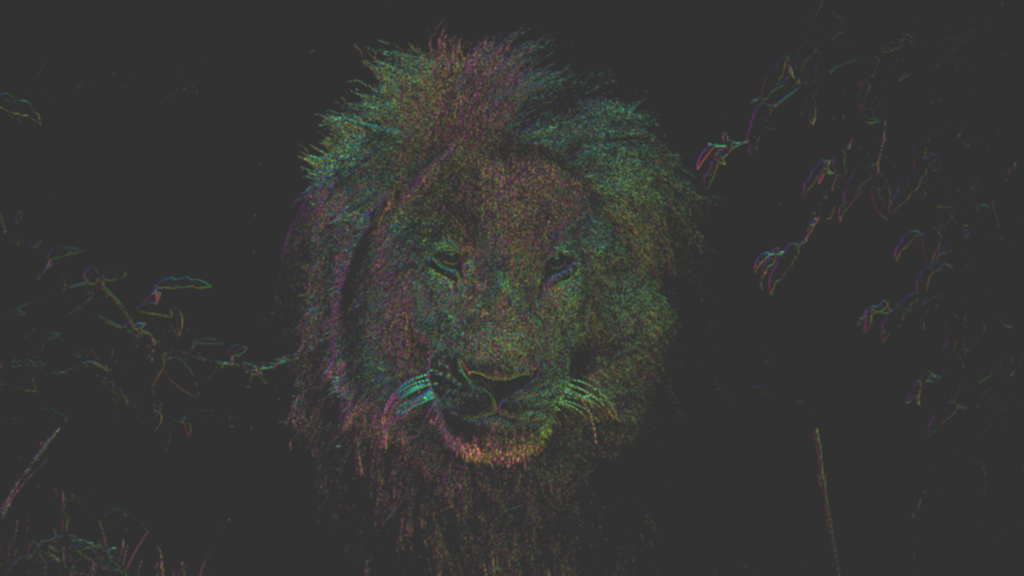}
    \caption{}
  \end{subfigure}
      \begin{subfigure}[b]{0.24\linewidth}
    \includegraphics[width=\linewidth]{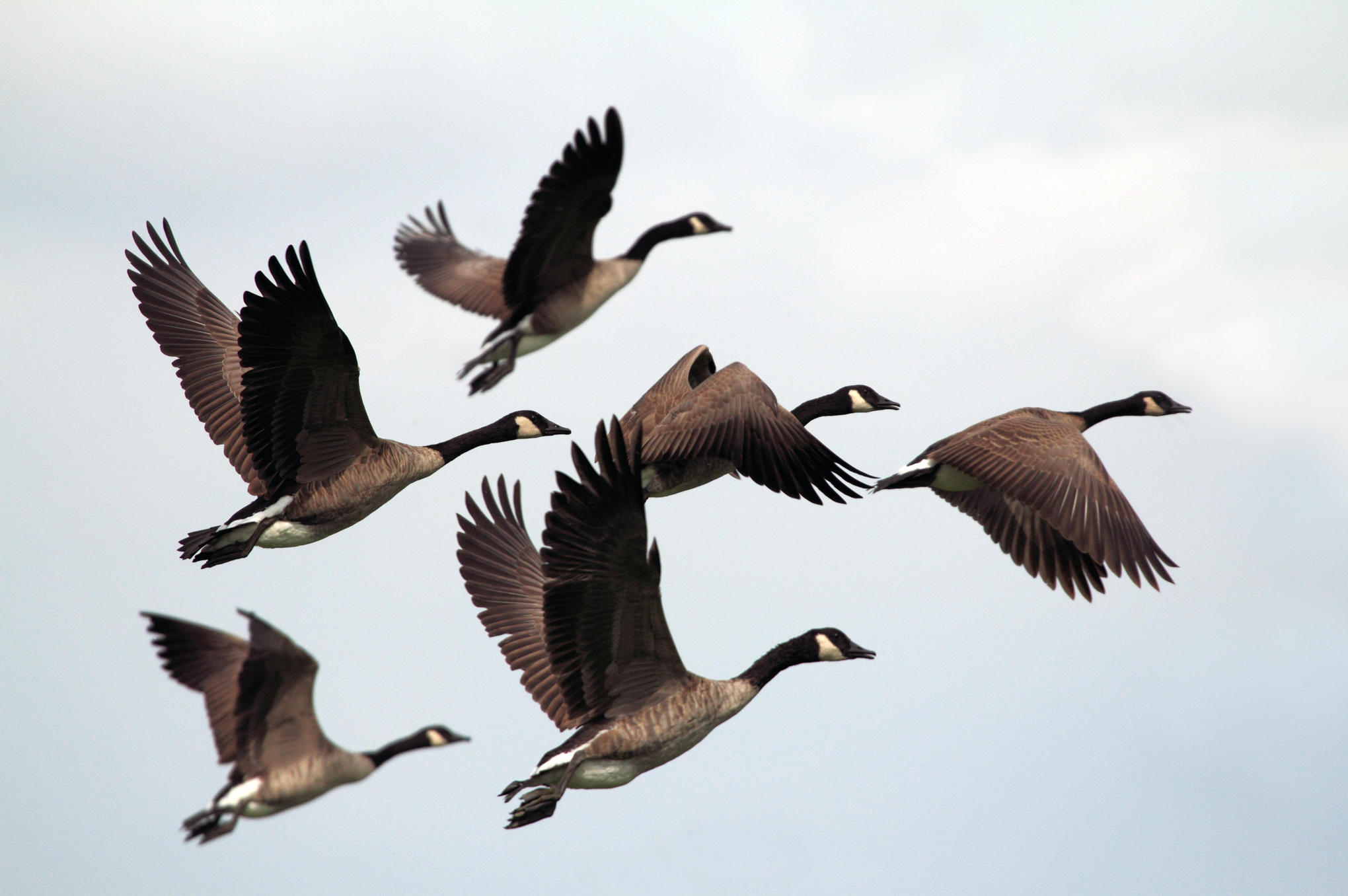}
     \caption{}
  \end{subfigure}
    \begin{subfigure}[b]{0.24\linewidth}
    \includegraphics[width=\linewidth]{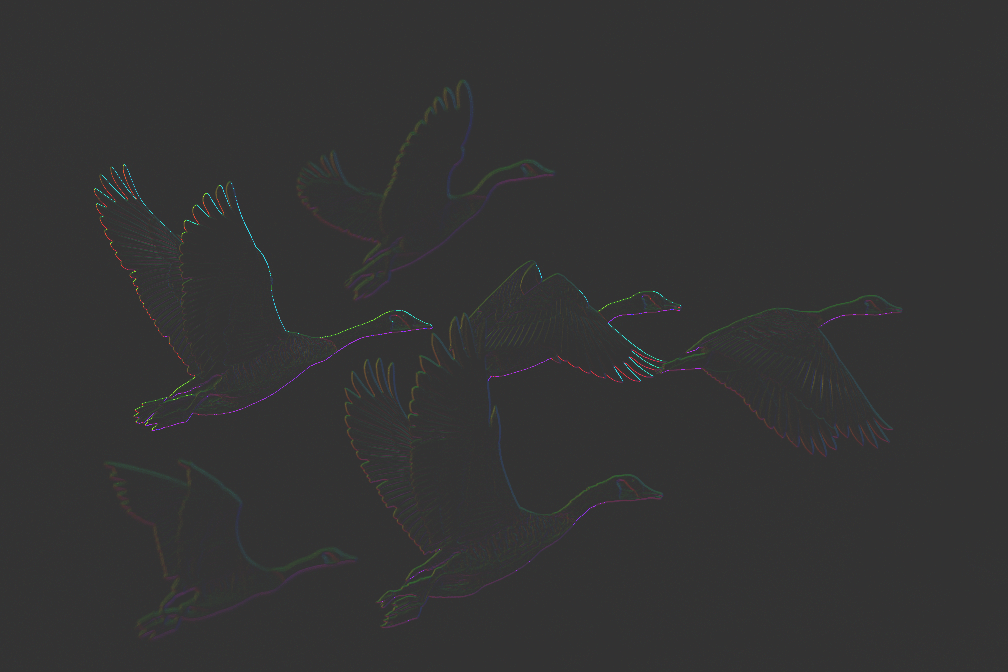}
    \caption{\textit{MAE}=2.57}
  \end{subfigure}
  \begin{subfigure}[b]{0.24\linewidth}
    \includegraphics[width=\linewidth]{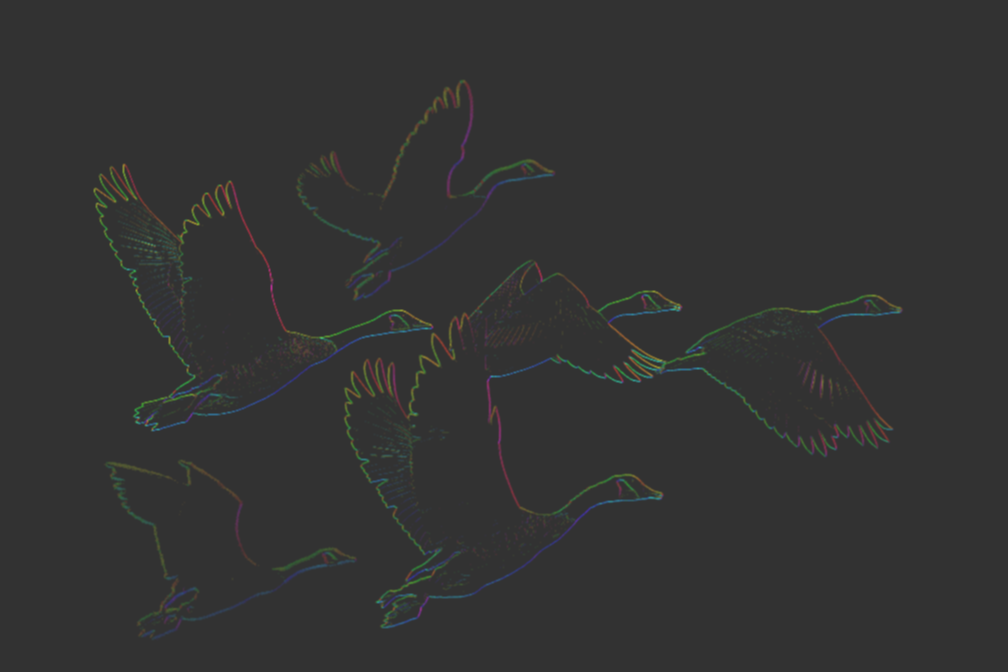}
    \caption{\textit{MAE}=1.34}
  \end{subfigure}
    \begin{subfigure}[b]{0.24\linewidth}
    \includegraphics[width=\linewidth]{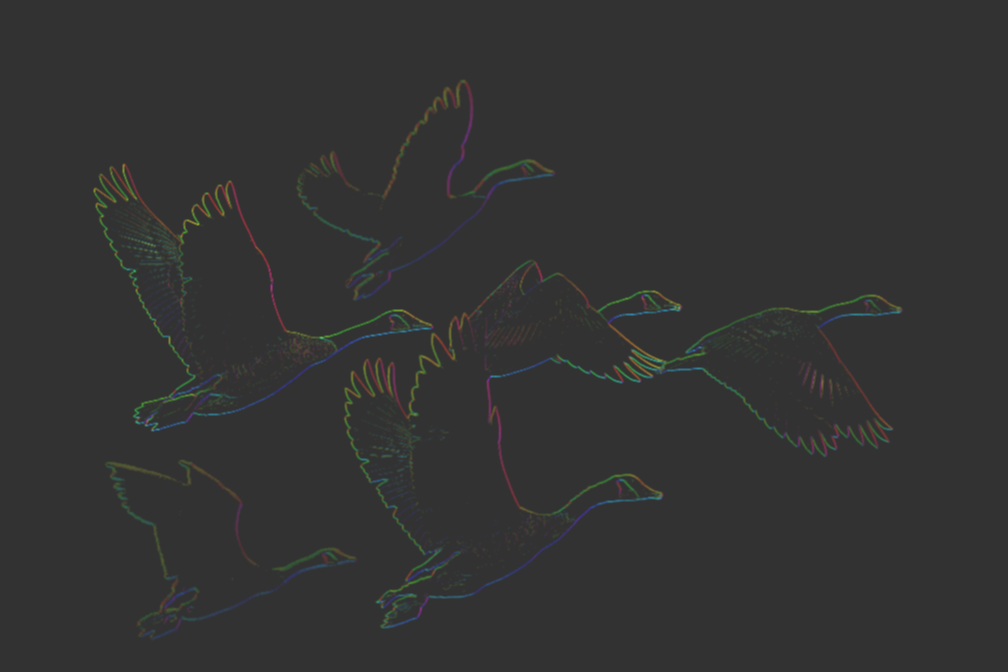}
    \caption{}
  \end{subfigure}
        \begin{subfigure}[b]{0.24\linewidth}
    \includegraphics[width=\linewidth]{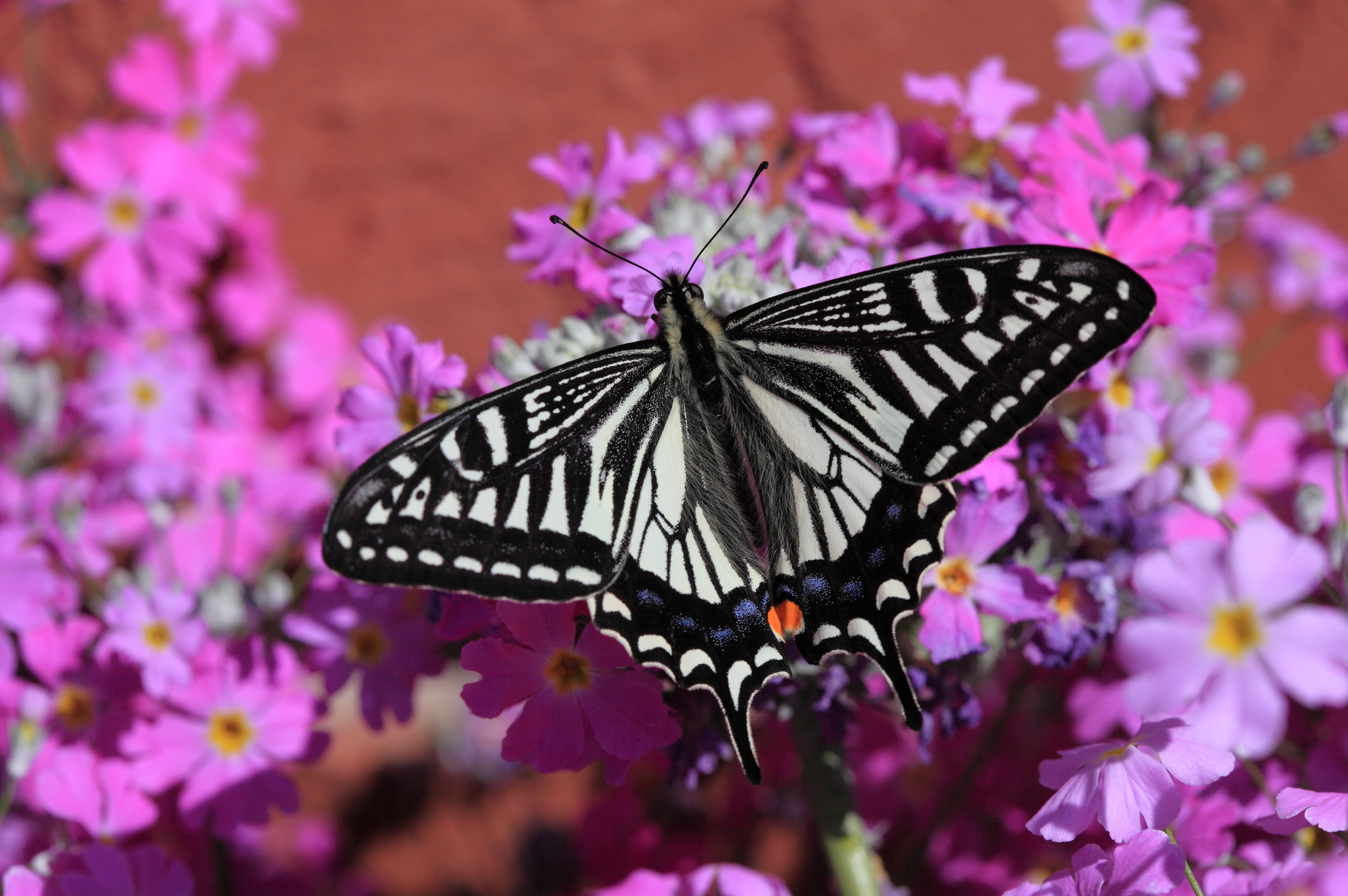}
     \caption{}
  \end{subfigure}
      \begin{subfigure}[b]{0.24\linewidth}
    \includegraphics[width=\linewidth]{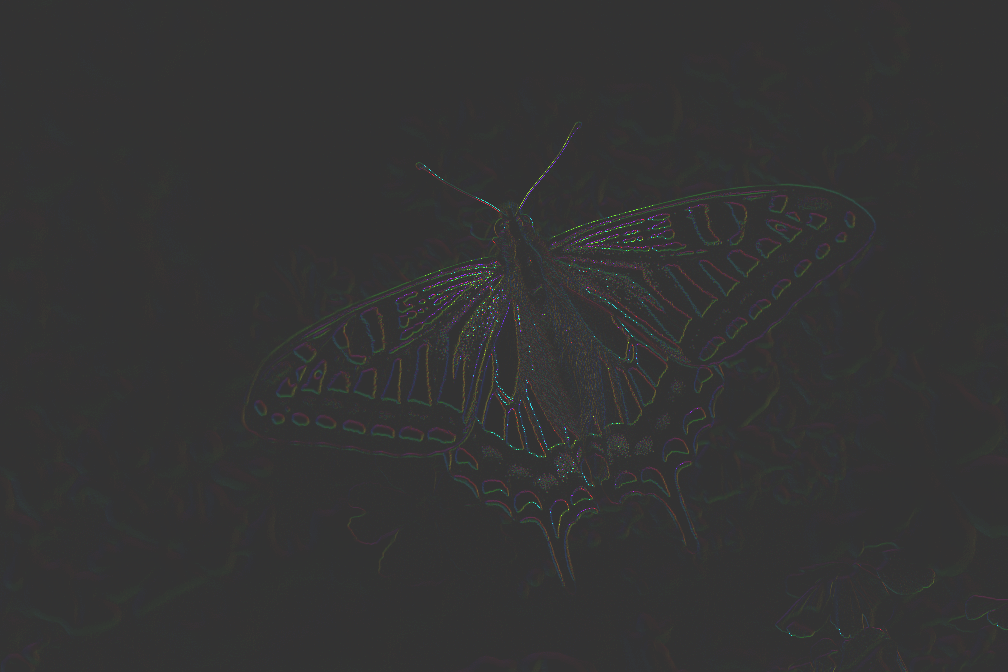}
    \caption{\textit{MAE}=3.21}
  \end{subfigure}
  \begin{subfigure}[b]{0.24\linewidth}
    \includegraphics[width=\linewidth]{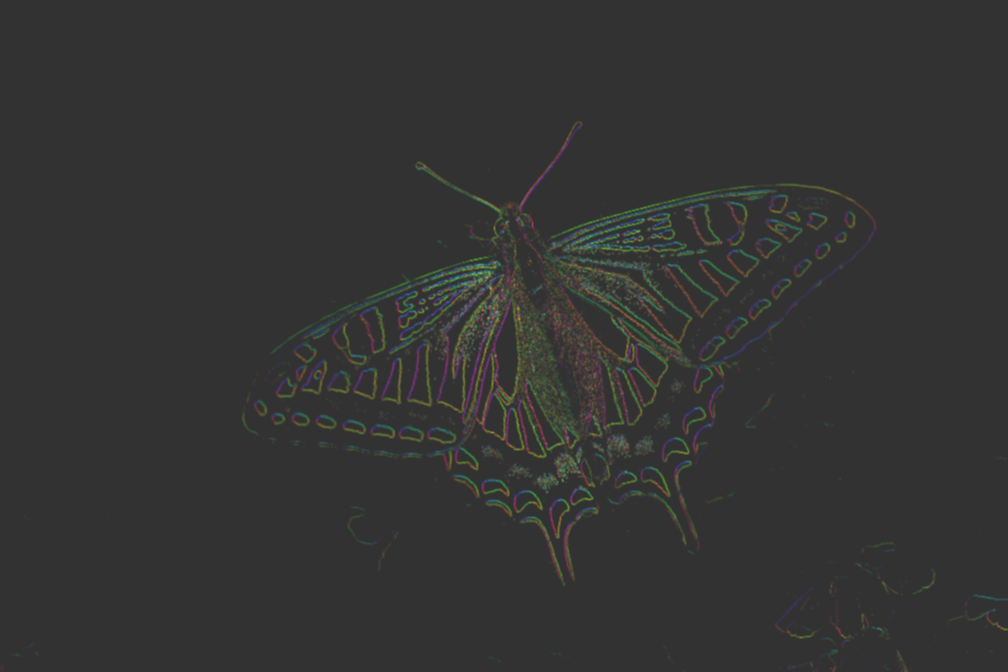}
    \caption{\textit{MAE}=1.36}
  \end{subfigure}
    \begin{subfigure}[b]{0.24\linewidth}
    \includegraphics[width=\linewidth]{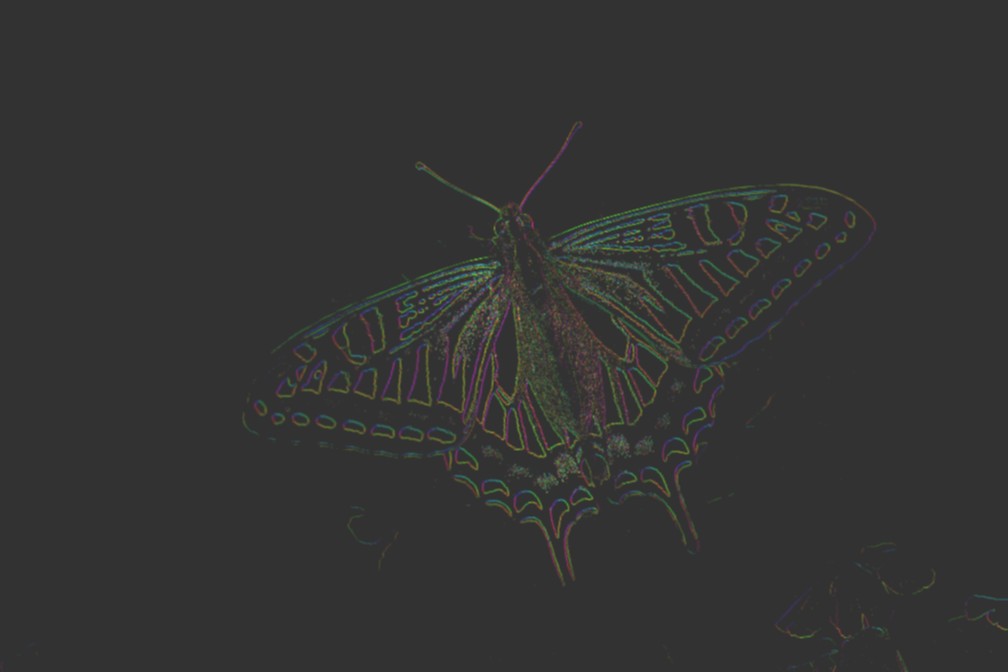}
    \caption{}
  \end{subfigure}
  \caption{\textbf{Visualizing the improved precision of quantized gradient images $I_\nabla$ by soft lattice dequantization (LQ$^{-1}$) module for some test samples.} From left to right: the original images, quantized gradient images, improved gradient images after soft dequantization LQ$^{-1}$, the true gradient images.}
  \label{fig:grad_vis}
\end{figure*}

To appreciate how effective the LQ$^{-1}$ soft dequantization is to reduce the quantization error, we present, in Figure \ref{fig:grad_vis}, four examples of
the lattice quantized gradient image $(Q(g),Q(h))$, the LQ$^{-1}$ restored gradient image $(\hat{g},\hat{h})$, and the original gradient image $(g,h)$, so one can visualize the accuracy improvements of LQ$^{-1}$ over hard dequantization in both angle (coded by color) and amplitude (coded by intensity) of the local gradients.  As an objective error metric, the mean absolute errors (MAE) for each sample image are also given in the Figure \ref{fig:grad_vis}.

\subsubsection{Different Upsampling Modules}
The proposed method can use any available super-resolution network as the upsampler as long as they are differentiable. We have tested our method with three well-known super-resolution networks: RCAN \cite{zhang2018rcan}, EDSR \cite{lim2017enhanced}, and WDSR \cite{yu2018wide}.
The results are reported in Table \ref{tab:resultsbackbone}.

\begin{table}
\centering
\caption{\label{tab:resultsbackbone} \textbf{Comparison of various super resolution networks as the upsampler module in SR-LQ$^{-1}$}.}
\label{tab:performance}
\begin{tabular}{cccc}
 & SR-LQ$^{-1}$(EDSR) & SR-LQ$^{-1}$(WDSR) & SR-LQ$^{-1}$(RCAN) \\
\hline
\textcolor{red}{\textbf{DIV2K}}   \\
PSNR & 41.82 & 41.49 & 41.53 \\
SSIM & 0.99 & 0.99 & 0.99 \\
[0.3cm]
\hline
\textcolor{red}{\textbf{BSD100}}  \\
PSNR & 39.00 & 38.83 & 38.90 \\
SSIM & 0.99 & 0.99  & 0.99  \\
[0.3cm]
\hline
\textcolor{red}{\textbf{Set14}} & & &  \\
PSNR & 39.61 & 39.43 & 39.57 \\
SSIM & 0.99 & 0.99 &  0.98 \\
[0.3cm]
\hline
\textcolor{red}{\textbf{Set5}} & & & \\
PSNR & 42.92 & 42.81 & 42.85 \\
SSIM & 0.99 & 0.98 &  0.99 \\
[0.3cm]
\hline
\end{tabular}
\end{table}

\subsubsection{Comparing various quantizers}
\begin{figure}[H]
  \centering
  \includegraphics[width=0.8\linewidth]{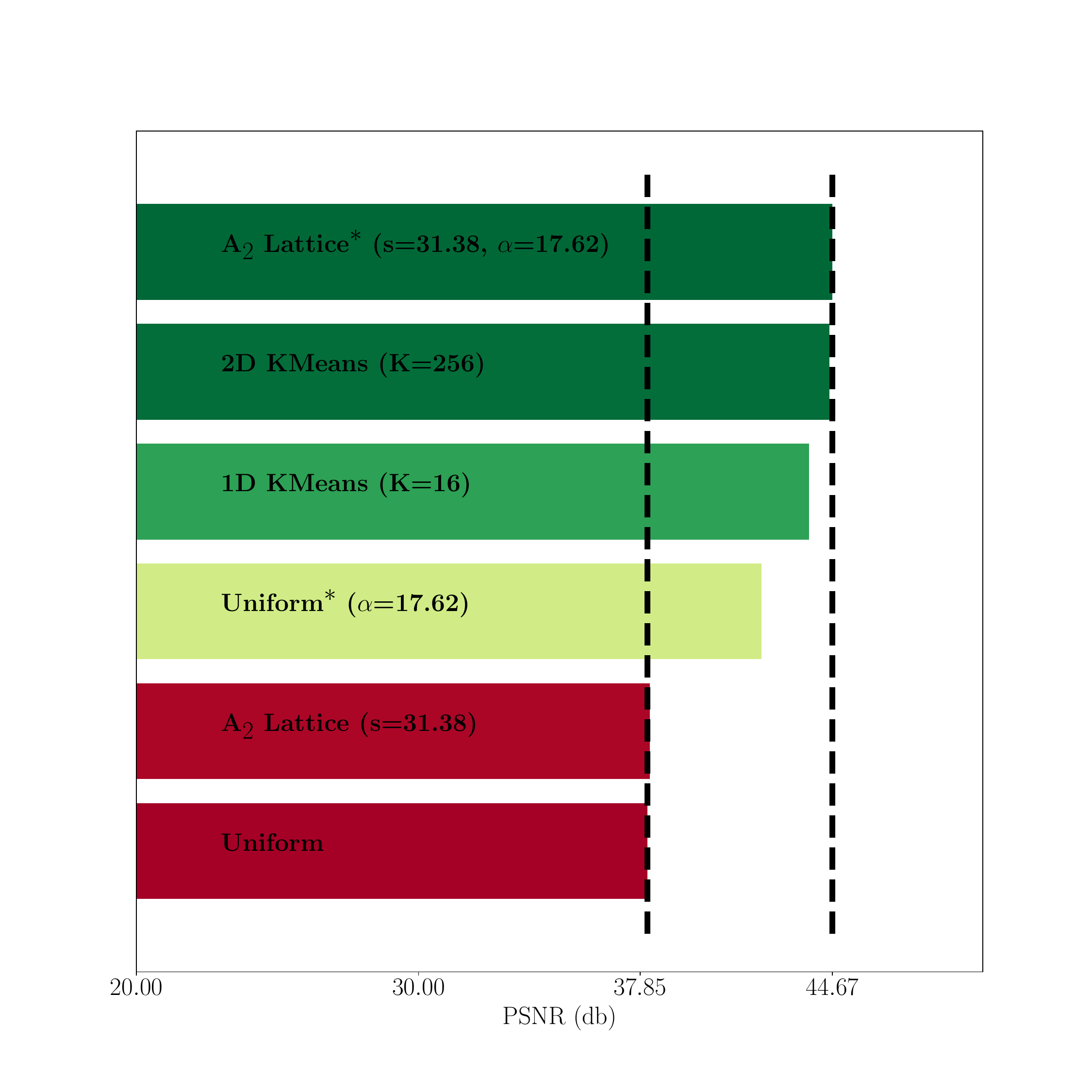}
  \caption{\textbf{PSNR of different quantizers}. $*$ means A-law companding is performed before quantization.}
  \label{fig:quantizers}
\end{figure}

To demonstrate the advantages of the $A_2$ lattice quantizer for encoding the gradient image $I_\nabla$ in the proposed ACUTE compression system, we compare it, in Figure \ref{fig:quantizers}, against various other quantizers in quantization precision (PSNR), including
1D Lloyd-Max quantizer (1D K-means), 2D Lloyd-Max quantizer (2D K-means), and uniform quantizer.  In the case of 1D K-means, we design two optimal scale quantizers for $g$ and $h$ separately with $K=16$.  In the case of 2D K-means, we design a 2D vector quantizer of $K=256$ codewords by clustering $g$ and $h$ in the 2D space.
As exhibited, the proposed lattice quantizer coupled with companding has the best performance in the comparison group.

\subsubsection{Do companding and $A_2$ quantization matter?}
\begin{figure}[H]
  \centering
  \includegraphics[width=0.9\linewidth]{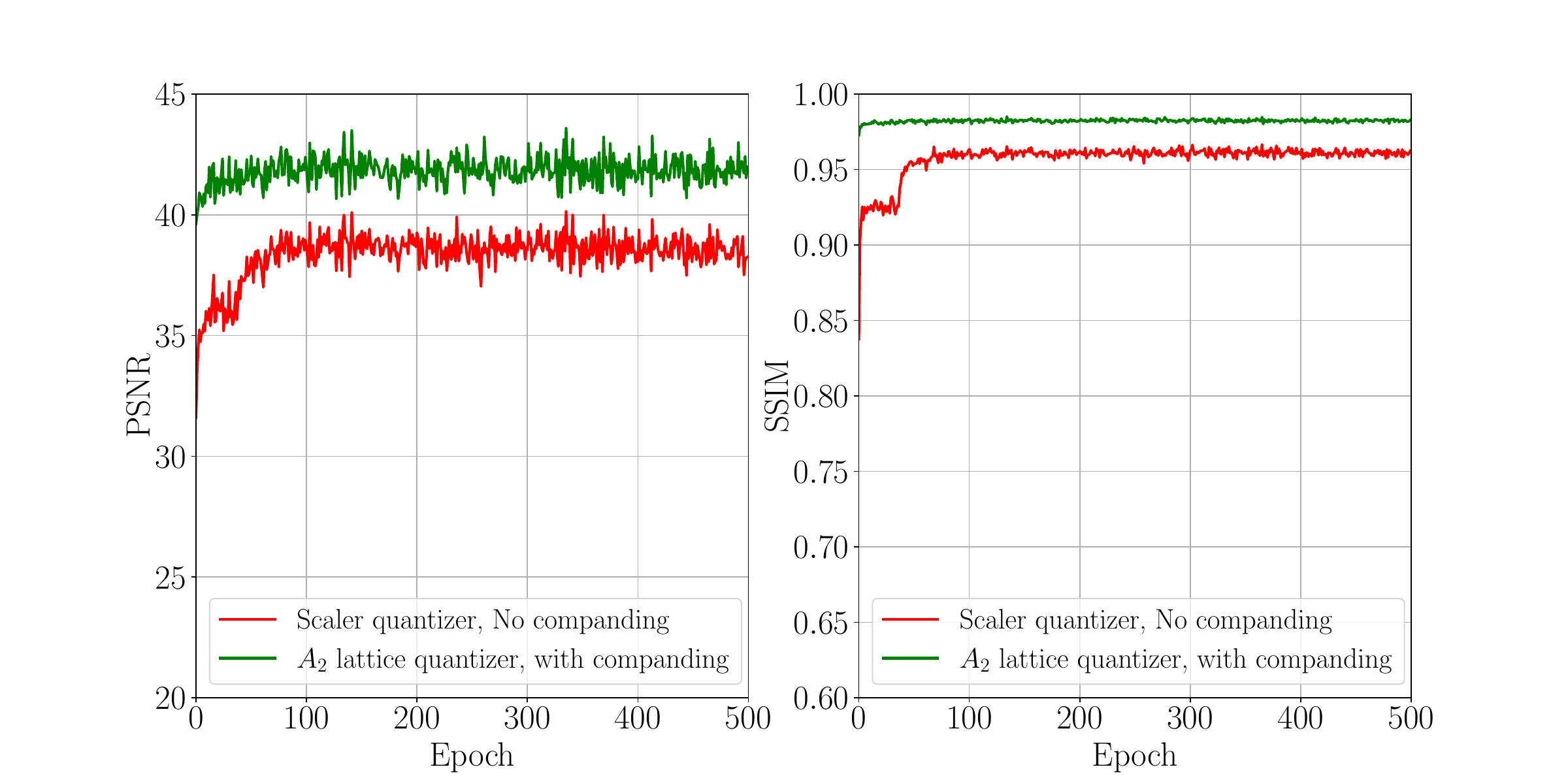}
  \caption{\textbf{PSNR and SSIM trends vs the number of training epochs for DIV2K validation data}.  The red graphs are for scalar qunatizer without companding.  The green graphs demonstrate the higher performance of SR-LQ$^{-1}$.}
  \label{fig:q2exp}
\end{figure}

One may question the necessity of employing the $A_2$ quantizer coupled with companding at the encoder, considering that
a powerful deep learning-based decoder like SR-LQ$^{-1}$ can be made highly nonlinear and complex to compensate for the rigidity of uniform quantization.  
To clear any doubt,
we train the deep decoder separately for the scheme of 
companding followed by $A_2$ 2D lattice quantization and for simple uniform scalar quantization of $z, g, h$.
PSNR and SSIM values of reconstructed images, averaged over the DIV2K validation data, are plotted and compared in Figure \ref{fig:q2exp} for the two schemes.  The same architecture for the SR module is used in the two cases to have a fair comparison.
These results show that the latter is considerably inferior to the former in reconstruction quality.


\section{Conclusion}
We have developed an asymmetric image compression system (ACUTE) of an ultra-high throughput encoder coupled with a powerful deep soft decoder.  The proposed ACUTE system can sustain long duration of real-time video compression at very high frame rates or similar scenarios, while offering competitive rate-distortion performance.
The light, hardware-friendly but efficient encoder is made possible by simple down sampling and 2D lattice quantization of gradients with companding.  The learnt SR-LQ$^{-1}$ decoder thoroughly exploits the gradient information transmitted by the encoder to superresolve the downsampled image in good precision.


\clearpage
\printbibliography

\end{document}